\documentclass[aps,pra]{revtex4-2}

\usepackage{graphicx}
\usepackage{amsmath,amssymb}
\usepackage{hyperref}

\begin{document} 

\title{A non-local exchange potential for electronic structure calculations}

\author{G. Schiwietz}%
\affiliation{Helmholtz-Zentrum Berlin für Materialien und Energie GmbH,  
 Albert-Einstein-Str. 15, 12489, Berlin, Germany 
}%

\author{P.L. Grande}
\affiliation{Instituto de F\'isica da Universidade Federal do Rio Grande do Sul, Av. Bento Gon\c{c}alves, Porto Alegre, 9500, Brazil
}%

\begin{abstract}
In this work we describe a model for the exchange interaction of electrons,
as it follows from the Pauli exclusion principle. Starting from Hartree-Fock
theory and making use of the free electron-gas model we propose a simple
scheme to calculate the exchange-potential for atoms as well as simple
molecules and solids. This method assures
the correct asymptotic long-range behavior of the potential, contrary to
local-density approximations that rely on a strongly simplified generalization 
of the well-known Kohn-Sham or Slater exchange interaction. 
Furthermore, our results approach the Kohn-Sham results in the
interstitial space of solids. As a benchmark test, 
total energies and eigenenergies for atoms from He to Xe 
computed within our model are compared to other calculations
as well as to experimental data.
\end{abstract}


\maketitle

\section{Introduction}
\label{sec:Introduction}

There are many problems in physics and also in quantum chemistry where the
electronic structure or the electronic motion in multi-electron systems have
to be treated. They all have in common that one generally has to consider
the mean electron--electron interaction, the antisymmetrization effects due
to the Pauli exclusion principle and, to some extend, the correlated
electronic motion. Most of the corresponding models are effectively
single-electron models that incorporate approximate descriptions of the
interaction potentials. Generally it is possible to treat the most important
contributions exactly, namely the electron--nucleus and the mean electron--electron interaction.
Correlation and the typically more important exchange terms, however, 
are often quantitatively not well described. Contrarily, in atomic-structure
calculations one has to deal with a finite number of electrons and 
with a single atomic center that also allows one to make use of the central-field
approximation. Thus, it is nowadays possible to solve the Hartree-Fock
equations \cite{Froese,Blume,Cowan}, the Kohn-Sham equations without
correlation \cite{Sharp,Talman,Krieger} or even the full relativistic
Dirac-Fock equations \cite{Grant,Mann} directly. In this case,
antisymmetrization and the corresponding electron-exchange potential is
fully accounted for.

In most other applications of quantum theory such as the electron structure
of large molecules, clusters or solids, however, severe approximations are
usually introduced for the exchange interaction that produce a 
considerable uncertainty \cite{Cowan,Callaway,Dimmock}. Only a few recent
calculations on structure and dynamics come close to an exact self-consistent-field (SCF) solution 
either without accurate correlation energies for solids \cite{Bylander,Gorling,Reeves,Maliyov} or 
with a limited basis set for time-consuming multi-configuration SCF 
calculations \cite{MCSCF}. 

Furthermore, the prediction of dynamical quantities, like photon- or charged-particle-induced transitions
probabilities, often also requires a unique potential for all electrons in
the initial {\em and} final (excited) states and thus again a simplification
of the exchange potential is needed. There are many works in which exchange
potentials have been investigated and reviewed \cite{Callaway,Dimmock,
xSlater,KohnSham,HermanCallaway,HermanDyke,Stoddart,Latter,
HermanSkillman,HedinL,Lundqvist,Barth,Katsnelson,Jones,Becke86, Becke88, PerdewA,
PerdewC,PerdewB} and the knowledge on the subject has reached a high level in the last decades.
As far as the prediction of the total energy is concerned, the generalized gradient 
approximations (GGAs) of Becke \cite{Becke86, Becke88} and 
of Perdew et al \cite{PerdewA,PerdewC,PerdewB} provide a particularly high precision. 
A detailed comparison of these expressions based on fits for the
exchange-energy density was published by Engel et al. \cite{Engel}. 
This motivates us to use Becke's original GGA \cite{Becke86} for comparison 
purposes. 
Nevertheless, there are some basic shortcomings in most approximations that
are currently applied to describe the exchange interaction and this is why
we think that there is still a need for improved potentials. 

In this work, we restrict ourselfes mainly to ab-inito full-potential methods 
that do not include fitted electron-exchange contributions. 
Further, we exclude exchange potentials for free monoenergetic particles 
which are important for the description of electron scattering \cite{Truhlar75,Truhlar80}. 
We also exclude effective-core \cite{Truhlar76} or pseudo-potential methods which are
specifically designed for describing one or a few orbital energies and 
asymptotic wavefunctions with high accuracy. Such solutions naturally 
exclude inner-shell orbitals which are vital for the description 
of x-ray absorption and emission as well as for Auger-electron production. 
Typically, the corresponding wavefunctions do not yield reliable 
electron-momentum distributions close to a nucleus. 
These momentum distributions, however, are 
important for charged-particle excitation, ionization and capture processes
in atomic collisions or ion-solid interactions.

What are now the requirements that define a ''good'' single-electron
potential for the corresponding representation of a multi-electron problem? 
Here we only state such conditions, by leaving out the problems connected with the
treatment of electron correlation, and more detailed explanations and discussions may
be found in sections \ref{subsec:HF}, \ref{subsec:FEG} and specifically \ref{subsec:NDX}:

\begin{description}
\item  [  (i)] Wavefunctions and orbital energies resulting from this potential 
as well as the corresponding orbital eigenenergies and total energies should be 
not too far from the exact Hartree-Fock solution.

\item  [ (ii)] Ground- and excited-state wavefunctions resulting from this potential 
should be orthogonal, e.g., for a simplified treatment of 
(non-linear) electron dynamics based on the independent-electron model.

\item  [(iii)] The potential should yield the correct asymptotic Coulombic behavior at large
distances from localized electron distributions.

\item  [(iv)] For a free electron gas, the average exchange potential should be
equal to the Slater result.

\item  [ (v)] The exchange potential acting on an electron at the Fermi level
should be close to the Kohn-Sham result.

\item  [(vi)] The potential should depend on the spatial electron density and not on
the wavefunctions, according to the Hohenberg-Kohn theorem \cite{Hohenberg}
and to simplify the implementation in existing programs.\label{conditions}
\end{description}

These conditions might seem contradictory, but in fact they are not 
as will be discussed in relation to the new electron-exchange potential.
It is emphasized there is no method that yields a unique potential for all
electrons as well as exact agreement with Hartree-Fock. Those potentials
that come close to most of the above requirements fail either for conditions
(v) and (vi) \cite{xSlater,HermanCallaway} or for (iv) and (v) \cite
{Sharp,Talman,Krieger,Gorling} and often there are significant
deviations from the Hartree-Fock orbital energies and binding energies. 

The potential that is proposed in this work fulfills the conditions (i) as well as 
(v) approximately and (ii), (iii), (iv) as well as (vi) exactly, unlike other ab-initio
potentials that are currently in use. After discussing the background of the
exact Hartree-Fock theory and implications from the free-electron-gas model,
this new potential is developed and presented in section \ref{subsec:NDX}. 
Experienced theorists that feel comfortable with the nomenclature 
used in this paper might directly start with section \ref{subsec:NDX}. 
Note, however, that the explanations on this potential and the 
corresponding results refer back to nearly 50\% of the equations 
in the preceeding sections \ref{subsec:HF} and \ref{subsec:FEG}. In section \ref{sec:Energies},
total energies and orbital energies calculated for atoms from He to Xe are
presented and compared to other calculations and also to experimental data.

\section{Approximate solutions of the Hartree-Fock method}
\label{sec:ApproximateSolutions}

Starting from Hartree-Fock theory we will discuss the basic properties of
the exchange potential with special emphasis on its asymptotic behavior. The
so-called exchange-charge density will be used in conjunction with results
for the free-electron-gas (FEG) model to develop an improved method (the
non-local-density exchange approximation, NDX) for the determination of the
exchange interaction from electron-charge densities for arbitrary elemental
atomic structures.

If not stated otherwise, atomic units (e = m$_e$ = $\hbar $ = 1 a.u.) will
be used throughout the paper. Thus, distances are measured in units of a$_0$
and energies in Hartree atomic units, corresponding to 27.2 eV.

\subsection{The Hartree-Fock theory}
\label{subsec:HF}

The quantum theory of electronic structure bases on the solution of the
stationary Schr\"{o}dinger equation, as given by

\begin{equation}
H(\vec{r}_1,\vec{r}_2,...,\vec{r}_N)\; \Psi _n(\vec{r}_1,\vec{r}_2,...,%
\vec{r}_N)\, =E_n\; \Psi _n(\vec{r}_1,\vec{r}_2,...,\vec{r%
}_N)\, ,  \label{Schr}
\end{equation}
with the total electronic energy $E_n$, the $N$-electron wavefunction 
$\Psi_n$ 
and the Hamiltonian $H$ that governs the free electron motion as well as
all interactions with each of the electrons at the coordinate vectors $\vec{r%
}_i$. The index $n$ stands for all quantum numbers necessary to describe a
specific state of the system. Eq. \ref{Schr} is applicable for a variety of
stationary systems such as atoms, molecule and clusters and also solids. The
nuclear coordinates are not explicitly included in the above Schr\"{o}dinger
equation, since they are assumed to be fixed in the adiabatic
Born-Oppenheimer approximation. From this point on, we will focus our
attention to one of the simplest electronic systems, namely isolated atoms.
The results of this work, however, are applicable also to more
complex electron systems.

Although seemingly simple, eq. \ref{Schr} cannot be solved analytically for
more than one electron. The non-relativistic Hamiltonian $H$ consists of a
sum over kinetic energy operators $T$, describing the motion of electrons,
and potential energy operators $V_{eff}$ that account for the interactions
of each electron with all other charged particles of the system. For
hydrogen-like atoms $V_{eff}$ reduces simply to $V^{n\,e}=-Z/r$, the Coulomb
potential for an electron in the field of a nuclear charge $Z$. For heavy
atoms one has to consider the nucleus--electron interactions $V^{n\,e}(\vec{r%
}_i)$ for each electron as well as the important electron--electron
interactions $V^{e\,e}(\vec{r}_i,\vec{r}_j)$ that couple the coordinates $i$
and $j$ of all electron pairs. The corresponding total Hamiltonian can be
decomposed in sums over single-electron operators and two-electron operators
in the following way: 
\begin{mathletters}
\begin{eqnarray}
H(\vec{r}_1,\vec{r}_2,...,\vec{r}_N)\, &=&\sum_{i=1}^NT(\vec{r}_i)+V_{eff,i}(%
\vec{r}_i)  \label{Ham1} \\
&=&\sum_{i=1}^N\left[ T(\vec{r}_i)+V^{n\,e}(\vec{r}_i)\right]
+\sum_{i<j}^NV_{i,\,j}^{e\,e}(\vec{r}_i,\vec{r}_j)  \label{Ham2} \\
&=&\sum_{i=1}^N\left[ \frac{-\Delta _{(i)}}2+\frac{-Z}{r_i}\right]
+\sum_{i<j}^N\frac 1{\left| \vec{r}_i-\vec{r}_j\right| }.  \label{Ham3}
\end{eqnarray}

The multi-configuration wavefunction $\Psi _n$ may be written as a sum over
individual Slater determinants $\Psi _n^C$ with the appropriate probability
amplitudes $a^C$ for each electron configuration $C$%
\end{mathletters}
\begin{mathletters}
\begin{eqnarray}
\left| \Psi _n(\vec{r}_1,\vec{r}_2,...,\vec{r}_N)\,\right\rangle
&=&\sum_Ca^C\; \Phi _n^C(\vec{r}_1,\vec{r}_2,...,\vec{r}%
_N) \label{wfn1} \\
&=&\sum_Ca^C\;\sum_p\frac{(-1)^p}{\sqrt{N\,!}}\;P\; \varphi _n^C(\vec{r%
}_1,\vec{r}_2,...,\vec{r}_N)  \label{wfn2} \\
&=&\sum_Ca^C\;\frac 1{\sqrt{N\,!}}\det \,\left| \varphi _i(\vec{r}%
_j) \right| ^C,\,  \label{wfn3}
\end{eqnarray}
where $P$ is the permutation operator and $p$ is the permutation index. The 
$ \varphi _n^C(\vec{r}_1,\vec{r}_2,...,\vec{r}_N) \,$ are
simple product functions of the type $ \varphi _\alpha (\vec{r}%
_1) \, \varphi _\beta (\vec{r}_2)\, ... \,%
 \varphi _\omega (\vec{r}_N) \,$ which are unique for a
certain configuration $C$ and the $ \varphi _i(\vec{r}_j) 
$ are orthogonal single-electron wavefunctions for each spin orbital $i$.
Here the Slater determinant or equivalently the permutation operator serves
to force the wavefunctions to obey the Pauli exclusion principle. Except for
the resulting so-called Pauli correlation, there is no electron--electron
correlation within a single configuration. But the linear combination of
configurations in eqs. \ref{wfn1} - \ref{wfn3} accounts for electron-correlation effects
due to the full electrostatic electron--electron interaction\- in eqs. \ref
{Ham2} and \ref{Ham3}.

The Hartree-Fock approximation consists now in neglecting all but one
configuration in eq. \ref{wfn1}. This means we decouple the correlated motion
of electron pairs and implicitly the $\left| \vec{r}_i-\vec{r}_j\right|
^{-1} $ terms in eq. \ref{Ham3} are replaced by the mean electron--electron
interactions $\left| \vec{r}_i-\vec{r}_j\right| _{mean}^{-1}$. Using now
this simplified independent-electron wavefunction together with eq. \ref
{Ham3} one may search for a stationary self-consistent solution. With the
help of the Lagrange variational ansatz to minimize the energy functional $%
E\left[ \Phi _n^C\right] =\left\langle \Phi _n^C\,\left| H\right| \Phi
_n^C\right\rangle /\left\langle \Phi _n^C\,|\,\Phi _n^C\right\rangle $ for a
certain configuration and consideration of the properties of the permutation
operator one finally arrives at a set of integro-differential equations for
each single-electron state \cite{Cowan,Jones,Messiah}. Following the work of
Slater \cite{xSlater}, these Hartree-Fock equations \cite{Fock} for each
orbital $\lambda =1\,..\,N$ with orbital energy $\varepsilon _\lambda $ may
be written as 
\end{mathletters}
\begin{equation}
\left[ \frac{-\Delta }2+\frac{-Z}r+V_{direct}^{e\,e}(\vec{r})+V_{x,\lambda
}^{e\,e}(\vec{r})\right] \;\varphi _\lambda (\vec{r})=\varepsilon _\lambda
\;\varphi _\lambda (\vec{r})  \label{HF}
\end{equation}
provided the basis functions $ \varphi _\lambda (\vec{r}) 
$\ are orthonormal{\bf , }i.e.{\bf , }$\left\langle \varphi _\lambda %
\mid \varphi _\mu \right\rangle =\delta _{\lambda ,\mu }$. It is
noted that the step-wise numerical solution of the above equations requires
additionally non-diagonal Lagrange multipliers to satisfy the orthogonality
condition, if at least one sub shell is open \cite{Hartree,Froese}. The
direct and the exchange potentials, $V_{direct,\lambda }^{e\,e}(\vec{r})$
and $V_{x,\lambda }^{e\,e}(\vec{r},\vec{r}^{\prime })$, in eq. \ref{HF} are 
\begin{mathletters}
\begin{eqnarray}
V_{direct}^{e\,e}(\vec{r}) &=&-\int d\vec{r}^{\prime }\,\frac{\rho (\vec{r}%
^{\prime })}{\left| \vec{r}-\vec{r}^{\prime }\right| }  \label{V1} \\
V_{x,\lambda }^{e\,e}(\vec{r}) &=&-\int d\vec{r}^{\prime }\frac{\rho
_\lambda (\vec{r},\vec{r}^{\prime })}{\left| \vec{r}-\vec{r}^{\prime
}\right| },  \label{V2}
\end{eqnarray}
where the direct potential represents the effect of the mean mutual
repulsion of electrons. The exchange potential is attractive and is due to a
localized density reduction of electrons with the same spin as the
''active'' electron. The charge and so-called exchange-charge densities, $%
\rho (\vec{r}^{\prime })$ and $\rho _\lambda (\vec{r},\vec{r}^{\prime })$,
in eqs. \ref{V1} and \ref{V2} are given by \cite{xSlater,Brauer,Streitwolf} 
\end{mathletters}
\begin{mathletters}
\begin{eqnarray}
\rho (\vec{r}) &=&-\sum_\mu \;\left| \varphi _\mu (\vec{r})\right| ^2
\label{rho-d} \\
\rho _\lambda (\vec{r},\vec{r}^{\prime }) &=&\frac{\varphi _\lambda (\vec{r}%
^{\prime })}{\varphi _\lambda (\vec{r})}\sum_\mu \;\varphi _\mu ^{*}(\vec{r}%
^{\prime })\,\varphi _\mu (\vec{r})\,\delta (m_{s,\mu },m_{s,\lambda }) 
\nonumber \\
&=&\sum_\mu \frac{\varphi _\lambda ^{*}(\vec{r})\,\varphi _\mu ^{*}(\vec{r}%
^{\prime })\,\varphi _\lambda (\vec{r}^{\prime })\,\varphi _\mu (\vec{r})}{%
\varphi _\lambda ^{*}(\vec{r})\,\varphi _\lambda (\vec{r})}\;\delta
(m_{s,\mu },m_{s,\lambda })  \label{rho-x}
\end{eqnarray}
or equivalently by a formulation where both sums $\sum_\mu $ are replaced by 
$\sum_{\mu \neq \lambda }$, since the diagonal terms $\mu =\lambda $ of the
charge and exchange-charge density are identical and thus, they cancel each
other, when inserted into the Hartree-Fock equations \ref{HF}. Without the
diagonal term $\mu =\lambda $, the direct electrostatic potential $%
V_{direct,\lambda }^{e\,e}(\vec{r})$ is due to the interaction of a each
(''active'') electron with the (smeared out) charge densities of all {\em %
other} electrons. If additionally the exchange contribution is set to zero,
eq. \ref{HF} would exactly give the Hartree SCF solution 
\cite{Hartree0}, the intuitive solution for distinguishable particles and
the correct solution for the He ground state. The Hartree result can also be
obtained from the above formulation by replacing $\rho _\lambda (\vec{r},%
\vec{r}^{\prime })$ with $\left| \varphi _\lambda (\vec{r})\right| ^2$, the
self-interaction of the ''active'' electron in the state $\lambda $.

The Kronecker symbol $\delta (m_{s,\mu },m_{s,\lambda })$ in eq. \ref{rho-x}
causes the summation to be carried out only over states of the same spin $%
m_{s,\lambda }$. Furthermore, the exchange-charge density $\rho _\lambda (%
\vec{r},\vec{r}^{\prime })$ depends on the state $\lambda $ of the
''active'' electron and it is not symmetric with respect to the coordinates $%
\vec{r}$ and $\vec{r}^{\prime }$. Thus, all the problems of the solution of
eq. \ref{HF} have been shifted into the troublesome exchange potential $%
V_{x,\lambda }^{e\,e}(\vec{r})$. Most theoretical treatments of electronic
structure start from this or similar formulations. It is emphasized,
however, the diagonal terms $\mu =\lambda $ in the exchange-charge density
lead to an artificial increase of the exchange potential. Thus,
uncertainties in an approximate description of the exchange term will
produce larger errors in the results. Such sources of errors will be
discussed in the subsequent paragraphs and a method to circumvent most of
these problems will be proposed in this work.

Some properties of the exchange potential and the underlying exchange-charge
density should be discussed at this point, since they provide the basis of
the present work. The total exchange-charge acting on the state $\lambda $
is 
\end{mathletters}
\begin{mathletters}
\begin{eqnarray}
\int d\vec{r}^{\prime }\,\rho _\lambda (\vec{r},\vec{r}^{\prime }) &=&\int d%
\vec{r}^{\prime }\,\sum_\mu \;\frac{\varphi _\lambda ^{*}(\vec{r})\,\varphi
_\mu ^{*}(\vec{r}^{\prime })\,\varphi _\lambda (\vec{r}^{\prime })\,\varphi
_\mu (\vec{r})}{\varphi _\lambda ^{*}(\vec{r})\,\varphi _\lambda (\vec{r})}%
\delta (m_{s,\mu },m_{s,\lambda })  \label{cond3a} \\
&=&\sum_\mu \;\frac{\,\varphi _\mu (\vec{r})}{\varphi _\lambda (\vec{r})}%
\delta (m_{s,\mu },m_{s,\lambda })\int d\vec{r}^{\prime }\,\varphi _\mu ^{*}(%
\vec{r}^{\prime })\,\varphi _\lambda (\vec{r}^{\prime })  \nonumber \\
&=&\sum_\mu \;\frac{\,\varphi _\mu (\vec{r})}{\varphi _\lambda (\vec{r})}%
\delta (m_{s,\mu },m_{s,\lambda })\;\delta (\mu ,\lambda )  \nonumber \\
&=&1,  \nonumber
\end{eqnarray}
independent of $\vec{r}$ and of $\lambda $. Note that the result depends on
the formulation of the Hartree-Fock equations and it would be $0$, if the
diagonal element $\mu =\lambda $ would not have been included in the above
sum. Thus, the charge of $+1$ is due to the hole that compensates the charge
of the ''active'' electron itself (often denoted self interaction), since it
is also included in the direct electron--electron interaction. It
immediately follows from eqs.  \ref{V2} and \ref{cond3a} for a finite extension of the
wavefunctions (imagine, e.g., bound states centered around $r=0$) that 
\begin{eqnarray}
\lim_{r\rightarrow \infty }V_{x,\lambda }^{e\,e}(\vec{r})
&=&-\lim_{r\rightarrow \infty }\int d\vec{r}^{\prime }\,\frac{\rho _\lambda (%
\vec{r},\vec{r}^{\prime })}{\left| \vec{r}-\vec{r}^{\prime }\right| }
\label{cond3b} \\
&\approx &-1/r\int d\vec{r}^{\prime }\,\rho _\lambda (\vec{r},\vec{r}%
^{\prime })  \nonumber \\
&=&-1/r.  \nonumber
\end{eqnarray}
To obtain exactly this behavior (condition (iii), mentioned in the
introduction), a corresponding correction was already introduced into early
Hartree-Fock-Slater calculations of atomic structure \cite{HermanSkillman}.
The procedure that is most often applied to improve the large-r behavior of
the exchange potential is known as Latter correction \cite{Latter} and for
neutral atoms it is simply given by 
\begin{equation}
V_{HFS-L}^{eff}(\vec{r})=\min \left[ \frac{Z}r+V_{direct}^{e\,e}(\vec{r}%
)+V_{x,Slater}^{e\,e},1/r\right] .  \label{Latt}
\end{equation}
Here the local Slater exchange potential $V_{x,Slater}^{e\,e}$ (see eq. \ref
{Slater} below), to be discussed in the next section, is replaced by $-1/r$
at large values of $r$. One may argue that the kink that shows up in this
potential is unphysical, but from a more general point of view the main
problem is the restriction of this recipe to the pure atomic case.

Another property of the exchange-charge density that will be used later is
the behavior of $\rho _\lambda (\vec{r},\vec{r}^{\prime })$ for $\vec{r}=%
\vec{r}^{\prime }$ : 
\end{mathletters}
\begin{mathletters}
\begin{eqnarray}
\lim_{\vec{r}\rightarrow \vec{r}^{\prime }}\,\rho _\lambda (\vec{r},\vec{r}%
^{\prime }) &=&\sum_\mu \frac{\varphi _\lambda ^{*}(\vec{r})\,\varphi _\mu
^{*}(\vec{r})\,\varphi _\lambda (\vec{r})\,\varphi _\mu (\vec{r})}{\varphi
_\lambda ^{*}(\vec{r})\,\varphi _\lambda (\vec{r})}\;\delta (m_{s,\mu
},m_{s,\lambda })\;  \label{r0-a} \\
&=&\sum_\mu \;\left| \varphi _\mu (\vec{r})\right| ^2\;\delta (m_{s,\mu
},m_{s,\lambda })  \label{r0-b} \\
&\cong &\frac 12\sum_\mu \;\left| \varphi _\mu (\vec{r})\right| ^2
\label{r0-c} \\
&=&\frac 12\rho (\vec{r}).  \label{r0-d}
\end{eqnarray}
This equation is exact, if each orbital is occupied by two electrons of
opposite spin. Otherwise, this result may be viewed as the average over both
spin orientations. This result tells us that the density of electrons of the
same spin as the ''active'' electron (50\% of all electrons) is zero close
to the position $\vec{r}$ of that electron. Thus, the Pauli exclusion
principle does not allow electrons of the equal spin to approach each other,
independent of their state \cite{xSlater}, and this effect is called Fermi
or exchange hole. It is noted that leaving out the diagonal term $\mu
=\lambda $ in eq. \ref{r0-a}, the above result reduces to a value of $\frac
12\rho (\vec{r})-\left| \varphi _\lambda (\vec{r})\right| ^2$ for doubly
occupied orbitals (a small change for the FEG but a significant effect for
localized electrons) \cite{Cowan}.

To conclude the discussion on the implications of the exact Hartree-Fock
theory, the strength of the different contributions to the total energy of
atoms are estimated from calculations and experimental data. Given in the
order of their importance for light atoms, the static interaction terms are:

\begin{itemize}
\item  The nucleus--electron interaction $\frac{-Z}r$: for inner shells it
completely dominates the binding energy and for outer shells it is as
important as the other interaction terms.

\item  The mean direct electron--electron interaction $V_{direct}^{e\,e}$:
it exactly cancels the nucleus-electron interaction far away from the
nucleus ($\frac{+Z}r$ seen by an external electron outside the valence-electron orbit).

\item  The mean electron--electron exchange interaction $V_x^{e\,e}$ for 
$Z>2$: its relative energy contribution is a few percent; its
influence increases with increasing distance from the nucleus and outside
the valence-electron orbit it approaches $\frac{-1}r$, as will be shown
later.

\item  The residual electron--electron interaction (correlation terms) for 
$Z<12$: this contribution is roughly proportional to the number
of bound electrons and it has a maximum relative contribution of only 1.5 \%
of the total energy for He(1s$^2\,^1$S).

\item  The relativistic mass--velocity and Darwin terms for $Z>12$: 
these two terms are proportional to Z$^4$ and they cancel each other to
a large extent. For the heaviest atoms, however, the relative contribution
of the sum of both terms is about 10\%.

\item  The finite nuclear size effects, spin--orbit interaction and the Lamb
shift: these corrections are always small compared to most of the other
terms. For the heaviest atoms, however, they exceed the correlation energy
by far.
\end{itemize}

This comparison shows that the exchange interaction yields the 3$^{rd}$
largest contribution to the total energy for most atoms, molecules or
solids. In the next section (section \ref{subsec:FEG}) the free electron gas is discussed. For this
simple electronic system, the first energy terms are effectively replaced by
zero and electron exchange yields the most important potential-energy
contribution to the total energy.

\subsection{The free electron-gas model and local potential approximations}
\label{subsec:FEG}

The free electron-gas (FEG) model is the simplest of all multi-electron
systems and often helps to simplify more complex and more realistic
problems. Especially the conduction-band properties of some metals, such as
Al, may be reasonably well described within this picture. The FEG is a
system without any boundary that includes only the mean interactions between
electrons. Interactions with the nuclei are assumed to be smeared out over
the infinite volume, so that they give rise only to an offset of the
energies $\varepsilon _\lambda $ in eq. \ref{HF}. Without the nuclei and
without boundaries there is translational invariance and thus, the ground
state possesses the same symmetry. Under these conditions, the mean direct
electron--electron interaction $V_{direct}^{e\,e}$ also reduces to a
constant value, independent of $\vec{r}$ and of $\lambda $, that cancels the
interactions with the nuclei and the corresponding model Hartree-Fock
equations are 
\end{mathletters}
\begin{equation}
\left[ \frac{-\Delta }2+V_{x,\lambda }^{e\,e}\right] \; \varphi
_\lambda (\vec{r}) =\varepsilon _\lambda ^{FEG}\; \varphi
_\lambda (\vec{r}) .  \label{HF-FEG}
\end{equation}
Thus, the only physical quantity that determines the actual system is the
mean electron density $\rho _0$ and its influence on $V_{x,\lambda }^{e\,e}$
and on the distribution of occupied states, i.e. on the Fermi level. It is
seen that the exchange potential depends on $\lambda $, which makes it
energy dependent and furthermore, the density-of-states is zero for the
top-most level. This problem is then removed by including electron
correlations in an interacting electron gas or in the well-known simple
Sommerfeld model of a free electron gas, where the exchange potential is
implicitly replaced by an energy-independent offset of the eigenenergies $%
\varepsilon _\lambda ^{FEG}$. The resulting solutions are plane waves for
the electron momentum $k$, and $\varepsilon _\lambda ^{FEG}$ is equal to $%
k^2/2$ plus a usually undetermined constant. In the ground state of the free
electron gas all states are occupied up to the Fermi momentum $k_F$.

A FEG in the ground state is build up of plane waves of the type $\frac 1{%
\sqrt{\Omega }}e^{i\vec{k}\vec{r}}$ and the corresponding exchange hole
within Hartree-Fock theory (eq. \ref{rho-x}) may be derived from 
\begin{equation}
\rho _\lambda ^{FEG}(\vec{r},\vec{r}^{\prime })=\frac 1{2\Omega }e^{-i\vec{k}%
_\lambda (\vec{r}-\vec{r}^{\prime })}\sum_\mu \;e^{i\vec{k}_\mu (\vec{r}-%
\vec{r}^{\prime })}  \label{rho-FEG}
\end{equation}
where the factor $1/2$ accounts for both spin directions. The resulting
exchange-charge density has the two general properties of an exchange hole
as given by eqs. \ref{cond3b} and \ref{r0-d}. It is, however, complex valued
(the corresponding exchange potential is still real valued) and thus, this
expression is not convenient for a qualitative discussion of the exchange
hole.

On a more general basis, Slater has introduced an average exchange-charge
density $\bar{\rho}(\vec{r},\vec{r}^{\prime })$ that is the sum over the
individual exchange-charge densities weighted with the fractional density of
the ''active'' electron in the state $\lambda $ according to \cite{xSlater} 
\begin{mathletters}
\begin{eqnarray}
\bar{\rho}(\vec{r},\vec{r}^{\prime }) &=&\sum_\lambda \frac{\left| \varphi
_\lambda (\vec{r})\right| ^2}{\rho (\vec{r})}\rho _\lambda (\vec{r},\vec{r}%
^{\prime })  \label{aveX1} \\
&=&\sum_\lambda \frac{\left| \varphi _\lambda (\vec{r})\right| ^2}{\rho (%
\vec{r})}\sum_\mu \frac{\varphi _\lambda ^{*}(\vec{r})\,\varphi _\mu ^{*}(%
\vec{r}^{\prime })\,\varphi _\lambda (\vec{r}^{\prime })\,\varphi _\mu (\vec{%
r})}{\varphi _\lambda ^{*}(\vec{r})\,\varphi _\lambda (\vec{r})}\;\delta
(m_{s,\mu },m_{s,\lambda })  \label{aveX2} \\
&=&\sum_\lambda \sum_\mu \frac{\varphi _\lambda ^{*}(\vec{r})\,\varphi _\mu
^{*}(\vec{r}^{\prime })\,\varphi _\lambda (\vec{r}^{\prime })\,\varphi _\mu (%
\vec{r})}{\rho (\vec{r})}\delta (m_{s,\mu },m_{s,\lambda })  \label{aveX3}
\end{eqnarray}
This exchange-charge density is now independent of the state of the
''active'' electron and when inserted into eq. \ref{V2} and into the
FEG-Hartree-Fock equation \ref{HF-FEG}, it yields a density-of-states
consistent with the Sommerfeld model. The averaged exchange-charge density
and the resulting exchange potential is already quite close to what one
searches for. It fulfills the conditions (ii), (iii) and (iv) mentioned in
the introduction exactly and it is expected to yield results that are not so
far from the full Hartree-Fock solution. From the investigation of Ge \cite
{HermanCallaway} and Zn \cite{Krieger} one may extract that the averaged
exchange contribution to the orbital energies resulting from eq. \ref{aveX3}
are typically about 10\% too low for the inner-shell electrons.
Furthermore, the binding energy of the valence electrons is
overestimated. One disadvantage of this averaging procedure is that it will
not approach the Kohn-Sham exchange potential for delocalized
conduction-band electrons inside solids. Furthermore, it still depends on
the knowledge of all single-electron wavefunctions of the system and thus,
it complicates the implementation in the nowadays widespread
density-functional treatments of electronic structure. This criticism holds
true also for current exact solutions of the Kohn-Sham exchange potential 
\cite{Talman,Krieger,Gorling}.

The above definition of the average exchange-charge density may be used to
simplify the expression $\rho _\lambda ^{FEG}(\vec{r},\vec{r}^{\prime })$
for the FEG case. With the abbreviations $z=k_F\,l,$ $k_F\,=(3\pi ^2\rho
_0)^{1/3}$ and $l=\left| \vec{r}-\vec{r}^{\prime }\right| ,$ eq. \ref{aveX3}
reads then \cite{WignerSeitz} 
\end{mathletters}
\begin{eqnarray}
\bar{\rho}_x^{FEG}(\vec{r},\vec{r}^{\prime }) &=&\frac 12\sum_\lambda
\sum_\mu \frac{\varphi _\lambda ^{*}(\vec{r})\,\varphi _\mu ^{*}(\vec{r}%
^{\prime })\,\varphi _\lambda (\vec{r}^{\prime })\,\varphi _\mu (\vec{r})}{%
\rho _0}  \nonumber  \label{rhom-FEG1} \\
&=&\frac{\rho _0}2\frac 92\left[ \frac{\sin z-z\cos z}{z^3}\right] ^2,
\label{rhom-FEG}
\end{eqnarray}
which may be easily interpreted and also used to derive a simple expression
for the exchange potential $V_x^{e\,e}(\vec{r})$. Note that this function is
spherically symmetric, i.e., it depends only on $l$.

\begin{figure}[htb!]
    \centering
    \includegraphics[width=0.4\linewidth]{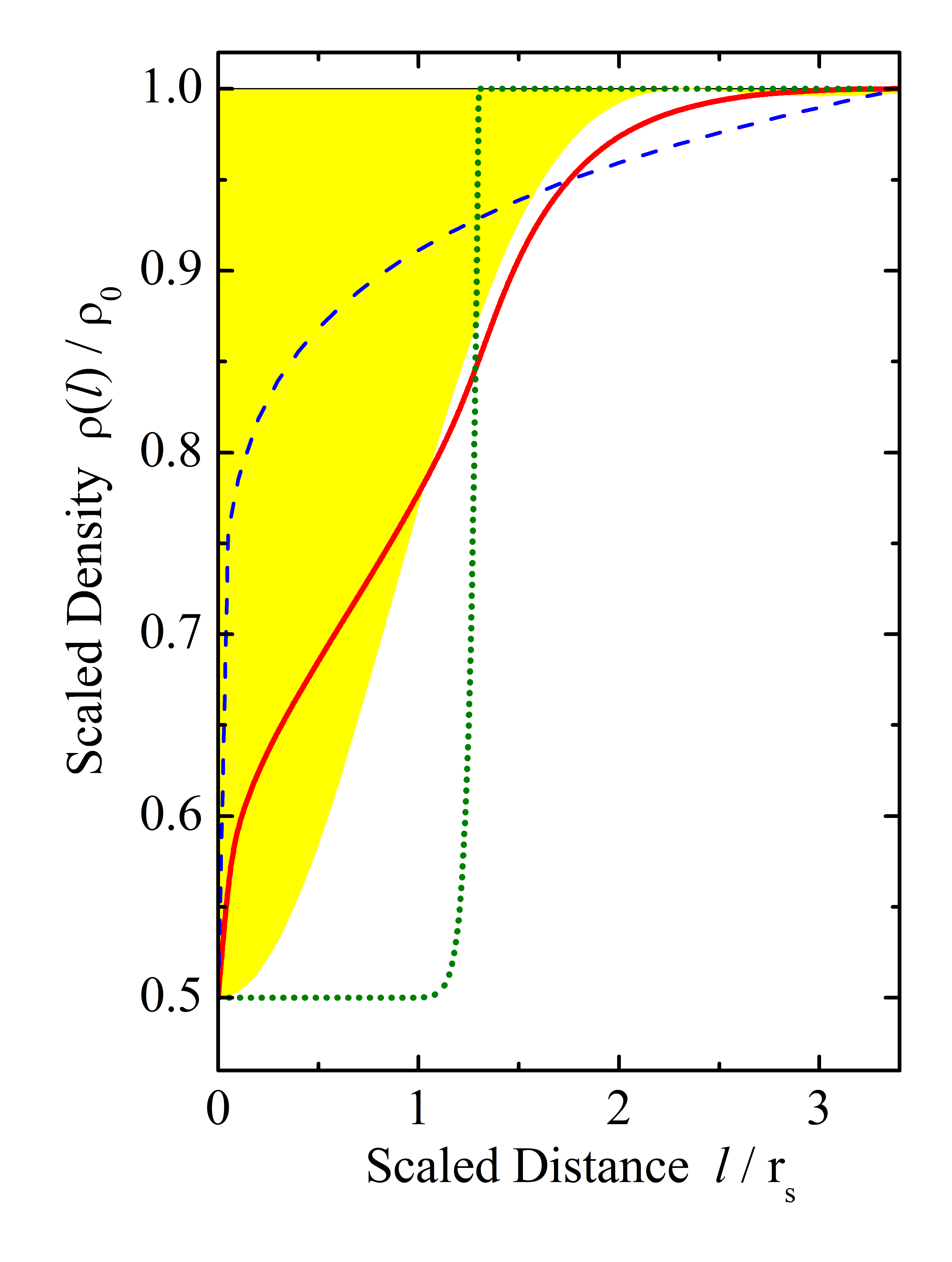}
    \caption{Distance dependent exchange-charge density. The yellow
area shows the shape of the exchange hole as it follows from the free
electron-gas picture. Our estimates for the shape of the exchange hole are
given by the dashed (corresponding to  large distances from the nucleus, $\alpha = 0.702$) and
the dotted curve (corresponding to distances close to the nucleus, $\alpha = 1.298$). The
averaged (over all distances from the nucleus) exchange-charge density of
the present NDX model is given by the solid red curve.}
    \label{fig:shape}
\end{figure}

Figure \ref{fig:shape} displays the normalized mean electron density (solid line
at the value 1) and the scaled exchange-charge density $1-\bar{\rho}%
_x^{FEG}(l)/\rho _0$ as a function of a scaled distance $l/r_s$, with the
scaling parameter $r_s$ being the mean radius 
\begin{equation}
r_s=(4\pi \rho _0/3)^{-1/3}  \label{r-s}
\end{equation}
of the sphere occupied per electron. The yellow area in the plot corresponds
to eq. \ref{rhom-FEG}. So, the effect of electron exchange via
antisymmetrization of the wavefunctions is a reduction of the electron
density in the close surrounding of any electron (Pauli correlation or
exchange hole) and the shape of the lower boundary of the yellow area gives
the effective electron density $\rho _0-\bar{\rho}_x^{FEG}(l)$ of {\em other}
electrons interacting with the ''active'' electron. The shape of this
exchange hole may be viewed as consisting of two parts: a) a large area that
exactly compensates the charge of the ''active'' electron, and b) an
oscillatory part, starting with a depletion at $l=0$, due to ''Pauli
repulsion'' of other electrons. The shape as plotted in Fig. \ref{fig:shape}%
, however, is completely determined by multi-electron effects, as can be
seen if eq. \ref{rho-x} is evaluated for an electron gas consisting only of
a two-electron singlet in a plane-wave state. In this case $\bar{\rho}%
_x^{FEG}$ would be independent of $l$ and it would cancel 50\% of $\rho $,
leaving a constant net charge of $\rho /2$.

At this point some comments about electron correlation appear to be
appropriate. Electron correlation is a two-electron effect that also
modifies the effective charge density near any ''active'' electron
(dependent on its state, but independent of its spin) and together with the
influence of exchange it gives rise to the so-called exchange-correlation
hole. The influence of correlation is expected to mainly reduce the electron
density close to the ''active'' electron down to about zero for low electron
densities (see, e.g., the density-functional results for antiprotons \cite
{Echenique}), since the electron--electron repulsion corresponds to a
barrier of infinite height at the origin $\vec{r}=\vec{r}^{\prime }$. It
should be emphasized that even for the simplest case of a FEG there are only
approximate analytical solutions of the full many-body Schr\"{o}dinger
equation including correlation. As may be seen from the comparison in the
last paragraph of section \ref{subsec:HF}, however, the all-over effect of
electron correlation is small enough to justify a treatment separately from
the exchange contribution.

In principle, an exchange potential for the free electron gas can be derived
from eq. \ref{rhom-FEG} by the use of eq. \ref{V2}. Such a potential is then
independent of $\vec{r}$. In his work, Slater \cite{xSlater} has proposed to
use this averaged FEG result 
\begin{equation}
V_{Slater}^{e\,e}(\vec{r})=-3\alpha \left[ \frac 3{8\pi }\rho (\vec{r}%
)\right] ^{1/3},{\rm with}\;\alpha =1  \label{Slater}
\end{equation}
and with the local electron density $\rho (\vec{r})$ (instead of $\rho _0$)
for realistic physical problems; and in fact, this expression has been and
it is still applied in many calculations of atomic or solid-state structure.
In an early work Dirac and afterwards also Kohn and Sham have derived
another expression for the exchange, using different methods. Their result,
often named Kohn-Sham exchange or 2/3-Slater exchange, corresponds to $%
\alpha =2/3$ in the above equation. This value is also representative for
the Fermi level of a free electron gas in the Hartree-Fock approximation. A
value of $\alpha =4/3$ may be derived for other the extreme, namely the
bottom level of an FEG \cite{Jones,Streitwolf}. It is noted that in atomic
calculations the value of $\alpha $ is sometimes varied (in the so-called X$%
_\alpha $ methods) to minimize the total energy functional or to fit total
Hartree-Fock energies. Such a fitting, however, is not attempted in this
work. The potential with $\alpha =1$ shows some significant deviations from
the HF results that cannot be explained by a constant factor \cite
{HermanCallaway}: at small distance from the nucleus it is roughly 15\% to
low and at larger distances (but still inside the outer shell) it might even
exceed the averaged Hartree-Fock results. Hence, the typical uncertainty of
simple exchange potentials is at least equal to the full correlation energy.

The above approximation to the exchange potential (eq. \ref{Slater})
corresponds to a replacement of the extended exchange hole in Fig. 
\ref{fig:shape} by a $\delta $-function at $\vec{r}=\vec{r}^{\prime }$ or
equivalently $l=0$. It is the (averaged) local-density approximation within
the Hartree-Fock theory of the FEG. This procedure is expected to be
accurate as long as no significant density variations occur within the
maximum boundary of the exchange hole. As can be seen from the figure,
however, this corresponds to a cut-off radius of $<2r_s $ ($<4$
to $11$ a.u. for the conduction-band density of metals). Since typical
lattice spacings have similar values and since density fluctuations due to
inner-shell electrons are much more localized, one may expect accuracy
problems for all methods that relate the exchange potential $V_x^{e\,e}(\vec{%
r})$ to the local density $\rho (\vec{r})$ only, e.g. the local Slater and
Dirac respectively Kohn-Sham methods, based on a generalization of the FEG
results.

It is emphasized that there are attempts to improve the local description of
the exchange potential by accounting for gradient terms in the charge
density, e.g. by Herman et al. \cite{HermanDyke} using fits to atomic
Hartree-Fock results or using a dimensional analysis by Stoddart et al. \cite
{Stoddart}. Later approaches by Becke \cite{Becke86,Becke88} and also by Perdew and coworkers are based on fits to
more accurate theoretical results and to experimental results for atoms \cite
{PerdewC} as well as di-atomic molecules \cite{PerdewB}. Although applied in
band-structure calculations \cite{WIEN}, all these results are not
satisfying from the stand point of an ab-initio solution of the problem.
Furthermore, it seems impossible to satisfy the asymptotic Coulomb condition defined by eq. \ref{cond3b}
with a local finite series expansion of the electron density (see also the discussion by Engel et al. \cite{Engel}). 
Thus, we expect a corresponding breakdown of the gradient corrections for strongly varying electron
densities (near the nuclei, at valence-electron node-structures and for low electron densities 
in the interstitial space of solids or molecules).

\subsection{The non-local-density exchange (NDX) approximation}
\label{subsec:NDX}

Summarizing the results of the preceding sections, an improved expression
for the exchange potential should better start from the description of an extended
exchange hole and not from a $\rho $-function representation as in the local
Slater or Kohn-Sham expressions of the exchange potential. Being guided by
eq. \ref{rhom-FEG} and to keep simplicity, the exchange hole should be real
valued. Solutions of the Schr\"{o}dinger equation with the corresponding
potential should be orthogonal. The easiest way to satisfy this condition is
the use of a unique energy-independent exchange hole for all single-electron
states, similar to the one in eq. \ref{aveX3}. From the first view, it seems
to be impossible to fulfill the requirements (iv) and (v) mentioned in the
introduction with an energy-independent potential. But let us postpone this
problem for a moment.

\begin{figure}[htb!]
    \centering
    \includegraphics[width=0.4\linewidth]{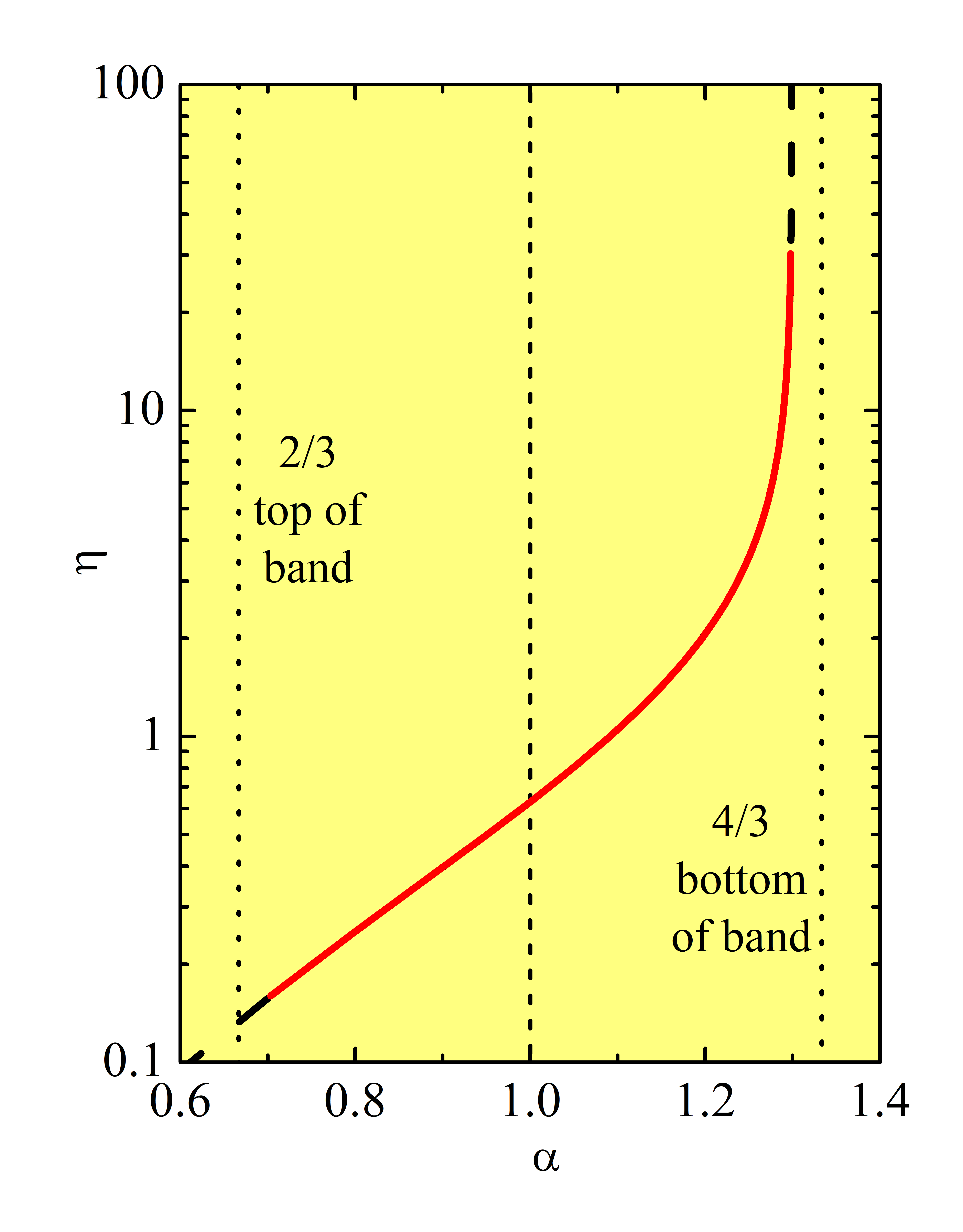}
    \caption{Exchange-hole parameter $\eta $
as a function of the dimensionless exchange strength $\alpha $.
The red part of the curve is used for the NDX exchange model.}
    \label{fig:eta}
\end{figure}

In order to simulate the properties of the exact exchange hole as good as possible we
also require eqs. \ref{cond3a}, \ref{cond3b} (condition (iii) on p. \pageref
{conditions}) and \ref{r0-d} to be true as constraints of the solution. From
the symmetry properties of eq. \ref{aveX3} it then follows that the
structure of $\bar{\rho}_x^{NDX}(\vec{r},\vec{r}^{\prime })$ is given by 
\begin{eqnarray}
\bar{\rho}_x^{NDX}(\vec{r},\vec{r}^{\prime }) &=&\frac{g(\vec{r},\vec{r}%
^{\prime })\;g(\vec{r}^{\prime },\vec{r})}{2\rho (\vec{r})}  \nonumber \\
&=&\frac{\left| g(\vec{r},\vec{r}^{\prime })\right| ^2}{2\rho (\vec{r})},
\label{G1}
\end{eqnarray}
where the function $g(\vec{r},\vec{r}^{\prime })$ should depend on the
density only and it is the complex conjugate of $g(\vec{r}^{\prime },\vec{r})
$. Thus, $\left| g(\vec{r}^{\prime },\vec{r})\right| ^2$ does not dependent
on the order of the variables and the function should also satisfy eq. \ref
{r0-d}. A very simple ansatz is then 
\begin{equation}
\left| g(\vec{r}^{\prime },\vec{r})\right| =\sqrt{\rho (\vec{r})\,\rho (\vec{%
r}^{\prime })\,G(\left| \vec{r}-\vec{r}^{\prime }\right| )}  \label{G2}
\end{equation}
and correspondingly with eq. \ref{G1} 
\begin{equation}
\bar{\rho}_x^{NDX}(\vec{r},\vec{r}^{\prime })=\frac{\rho (\vec{r}^{\prime })}%
2\,G(\left| \vec{r}-\vec{r}^{\prime }\right| ),  \label{G3}
\end{equation}
where some open problems have now been shifted into the determination of a
suitable function $G(l)$ with $l=\left| \vec{r}-\vec{r}^{\prime }\right| $.
It should be recalled, however, that the precise determination of its shape
is not so important, since already the introduction of an extended exchange
hole is expected to yield an improvement over the local Slater and Kohn-Sham
exchange methods, even when applying gradient corrections. $G(l)$ must be
integrable, in connection with eq. \ref{cond3a} and it should yield a value
of $1$ for $l=0$ (see eq. \ref{r0-d}). Furthermore, a simple but flexible
form of $G(l)$ should be chosen to allow for fast numerical integrations and
to fulfill additional constraints. One may of course argue that a natural
choice would be the shape of the FEG exchange hole (eq. \ref{rhom-FEG}), but
this function does not enable direct numerical integrations, since for an
inhomogeneous electron density it depends on $\rho (\vec{r}^{\,\prime })$ in
a complicated way. Thus, we have selected a very simple shape of the
exchange hole and by using the function $G(l)=1-\left( l/r_c\right) ^\eta $,
restricted to the range of positive values of $G$, we arrive at 
\begin{equation}
\bar{\rho}_x^{NDX}(\vec{r},\vec{r}^{\prime })=\frac{\rho (\vec{r}^{\prime })}%
2\max \left[ 1-\left( \left| \vec{r}-\vec{r}^{\prime }\right| /r_c\right)
^\eta ,0\right] .  \label{G4}
\end{equation}
Because we do not intend to perform any fittings of the exchange term, we
have not tried to vary the general form of eq. \ref{G4}. For the special case of a {\em constant}
electron density $\rho _0$, the two shape parameters, namely the cut-off
radius $r_c>0$ and the exponent $\eta >0$ are fixed by applying the general
condition of eq. \ref{cond3a} 
\begin{mathletters}
\begin{eqnarray}
1 &=&\frac{\rho _0}2\int d\vec{l}\,\max \left[ 1-\left( l/r_c\right) ^\eta
,0\right]   \nonumber \\
&=&2\pi \rho _0\int_0^{r_c}dl\;l^2-l^{\eta +2}/r_c^\eta   \nonumber \\
&=&2\pi \rho _0r_c^3\frac \eta {3(\eta +3)}  \label{G5a}
\end{eqnarray}
leading to 
\begin{equation}
r_c=\left( \frac{3\left( \eta +3\right) }{2\pi \rho _0\eta }\right) ^{1/3}.
\label{G5b}
\end{equation}
Furthermore, we require that the resulting exchange potential (using eq. \ref
{V2} and \ref{G4}) will match the FEG result in eq. \ref{Slater} 
\end{mathletters}
\begin{eqnarray}
-3\alpha \,\left[ \frac 3{8\pi }\rho _0\right] ^{1/3} &=&V_{x,FEG}^{e\,e} 
\nonumber \\
&=&-\frac{\rho _0}2\int d\vec{l}\;\frac{\max \left[ 1-\left( l/r_c\right)
^\eta ,0\right] }l  \nonumber \\
&=&-\pi \rho _0r_c^2\frac \eta {\eta +2}.  \label{G6}
\end{eqnarray}
From eqs. \ref{G5b} and \ref{G6} we now obtain $\eta $ as 
\begin{equation}
\alpha =\left( \frac{2\pi ^2\eta \left( \eta +3\right) ^2}{9\left( \eta
+2\right) ^3}\right) ^{1/3},  \label{G7}
\end{equation}
which may be solved once and forever, e.g., recursively or by tabulating $%
\alpha (\eta )$ and inverting the numerical relation. The computed dependence of 
$\eta(\alpha )$ is shown in Fig. \ref{fig:eta}. 

This relation for $\eta $ has the great advantage of being independent of $\rho _0$ 
as well as of $r_s$. Thus, one can assume that eq. \ref{G7} is also a good
approximation to the case of an inhomogeneous electron density $\rho (\vec{r}%
^{\prime })$. Here, the generalization to an inhomogeneous electron system
relies on the assumption that the inner part of the exchange-hole is
independent of any density variations. This is different from the
local-density assumption of Slater. 

Once, the parameter $\eta $ has been determined, one needs to calculate the
second shape parameter $r_c$. Generally, $r_c$ cannot be predicted prior to
the calculation, since for an inhomogeneous electron density $\rho (\vec{r}%
^{\,\prime })$ it depends on $\vec{r}^{\prime }$. The solution to this
problem is: one first solves eq. \ref{cond3a} through 
\begin{eqnarray}
1 &=&\frac 12\int_{l\leq r_c}d\vec{l}\;\rho \left( \vec{r}+\vec{l}\right)
\,\left[ 1-\left( l/r_c\right) ^\eta \right]   \nonumber \\
&=&\frac 12\int_{l\leq r_c}d\vec{l}\;\rho \left( \vec{r}+\vec{l}\right)
\,-\frac 1{2r_c^\eta }\int_{l<r_c}d\vec{l}\;l^\eta \,\rho \left( \vec{r}+%
\vec{l}\right)   \label{G8}
\end{eqnarray}
to determine $r_c$, the cut-off radius of the exchange hole in the direct
surrounding of the point $\vec{r}$. Then one may compute the exchange
potential at this point by the 3-dimensional integral 
\begin{eqnarray}
V_{x,NDX}^{e\,e} &=&\frac{-1}2\int_{l\leq r_c}d\vec{l}\;\rho \left( \vec{r}+%
\vec{l}\right) \,\frac{1-\left( l/r_c\right) ^\eta }l  \nonumber \\
&=&\frac{-1}2\int_{l\leq r_c}d\vec{l}\;\rho \left( \vec{r}+\vec{l}\right)
/l\,+\frac 1{2r_c^\eta }\int_{l\leq r_c}d\vec{l}\;l^{\eta -1}\,\rho \left( 
\vec{r}+\vec{l}\right) ,  \label{G9}
\end{eqnarray}
which reduces to a two-dimensional integral for the spherical symmetric
electron density in atoms. In order to save computing time one may also
integrate the four terms proportional to 
$\rho (\vec{r}+\vec{l})$, $l^\eta\,\rho (\vec{r}+\vec{l})$, 
$l^{-1}\rho (\vec{r}+\vec{l})$ and $l^{\eta -1}\rho (\vec{r}+\vec{l})$  
in eqs. \ref{G8} and \ref{G9} simultaneously. In a
step-wise integration, starting from $l=0$, one may use an automatic
step-width determination and stop both integrations exactly at the value of $%
l$ where the condition of eq. \ref{G8} is fulfilled ($l=r_c$). This is also
the method that was applied in this work. If only a reasonable estimate of
the potential and of the electron orbital energies are needed then one might
carefully select only a few points for the integration mesh. In this work,
however, we have optimized the accuracy of the numerical procedures to
compute total energies that are accurate to within about 6 digits, including
all correction terms.

Now, if we select $\alpha =1$ ($\eta =0.629$) the potential, defined by eqs. 
\ref{G7}, \ref{G8} and \ref{G9}, would satisfy the conditions (ii), (iii),
(iv) and (vi), but it would definitely deviate by about a factor of $3/2$
from the Kohn-Sham exchange expression. As discussed in section \ref{subsec:FEG},
the averaged exchange term (eq. \ref{aveX3}) shows significant deviations
from the full Hartree-Fock solutions and these deviations are nearly doubled
in the local free electron exchange by Slater (eq. \ref{Slater}). Contrary,
when the averaging procedure in eq. \ref{aveX3} is performed for each atomic
partial wave separately the accuracy is significantly improved \cite{HermanCallaway}. 
This means the less state selective the potential is the larger will be the deviation from
Hartree-Fock. Since our potential, however, should depend on $\vec{r}$ only,
we can only account for our knowledge on the type of states that occupy a certain range of $r$
(related to the distance from neighboring nuclei).

In any system consisting of electrons and nuclei, deeply bound states occupy small radii 
(small values of $r$ with respect to the neighboring nucleus) and weakly bound electrons, 
such as valence electrons or conduction-band electrons, have large radii. Recalling
the discussion on $\alpha $ (directly after eq. \ref{Slater}), it follows
that $\alpha $ should vary from a value of about $4/3$ for small $r$ to $2/3$
at large distances from the nuclei. In this work, we simply use the
following linear scaling for $\alpha $ 
\begin{equation}
\alpha (\vec{r}) = 1.298 - 2\cdot 0.298\cdot f^{far} (r) \label{G10}
\end{equation}
with the defining function $f^{far}$ that is zero close to the nucleus (for r=0) and 
reaches the value of 1 for electron coordinates far away from the nucleus. This function 
is given by the fractional electron charge $Q_r/Z$ inside a sphere of radius $r$ 
around the nucleus
\begin{equation}
f^{far} (r) = Q_r(r)/Z =\int_{r \prime \, \leq  \, r}d\vec{r} \, \prime \;
\rho \left( \vec{r} \, \prime \right) / Z .  \label{G11}
\end{equation}

Eq. \ref{G10} has been chosen in such a way that an average over the density
is always equal to one ($\bar{\alpha}=1$) independent of the density
distribution. This means it is consistent with condition (iv) of the introduction. 
Note that a maximum of $\alpha =4/3$ is in contradiction with
the selected shape of the function $G(l)$, see Fig. \ref{fig:eta}. Thus, we have
chosen a maximum value of $1.298$ and correspondingly the minimum value of $%
\alpha $ is $0.702$. Since valence electrons or conduction electrons test
mainly the regions far away from the nuclei, the effective value of $\alpha $
for these states will be about $0.7$ and thus, close to the Kohn-Sham
value of $2/3$ (see condition (v) on p. \pageref{conditions}). 

With eqs. \ref{G10} and \ref{G11} for $\alpha$, eq. \ref{G7} for $\eta$ 
as well as eqs. \ref{G8} and \ref{G9} for $r_c$ and $V_{x,NDX}^{e\,e}$ we have now
defined our approximation to the {\bf n}on-local-{\bf d}ensity e{\bf x}%
change (NDX) potential. This model is strictly mathematically also a local
potential method, since it finally depends only on the variable $\vec{r}$,
but through the non-local density contributions it incorporates the main
properties of the non-local exchange term as they follow from Hartree-Fock
theory. It is noted that the extension of the model to a spin-polarized
potential is straight forward: in eqs. \ref{G8} and \ref{G9} one simply has
to replace $\rho (\vec{r}+\vec{l})/2$ by $\rho _{\parallel }(\vec{r}+\vec{l})
$, the density of electrons with parallel spin.

Before the potential itself and the resulting eigenenergies and total
energies are discussed one may compare the analytical results for the shape
of the exchange hole (defined by eq. \ref{G4}) with the exact free electron
gas result in Fig. \ref{fig:shape}. For $\alpha =0.702$ ($\eta =0.1591$) the
scaled NDX exchange-charge density is plotted as a dashed blue curve. Its
magnitude is steeply decreasing and a large volume is tested by an exchange
hole of this shape. Contrary, for $\alpha =1.298$ ($\eta =30.47$) the
exchange hole has nearly a rectangular shape (dotted green curve) and reduces
the electron density only inside a smaller volume. The averaged (over all $%
\alpha $) exchange-charge density of the present NDX model is given by the
red solid curve under the assumption of a constant electron density. As can
be seen from Fig. \ref{fig:shape}, there is remarkable similarity of the 
NDX exchange hole with the exact averaged Hartree-Fock FEG result (yellow area), 
especially at distances $l/r_s>1$.

\subsection{Atomic Exchange potentials}
\label{subsec:NDXpot}

\begin{figure}[htb!]
    \centering
    \includegraphics[width=0.55\linewidth]{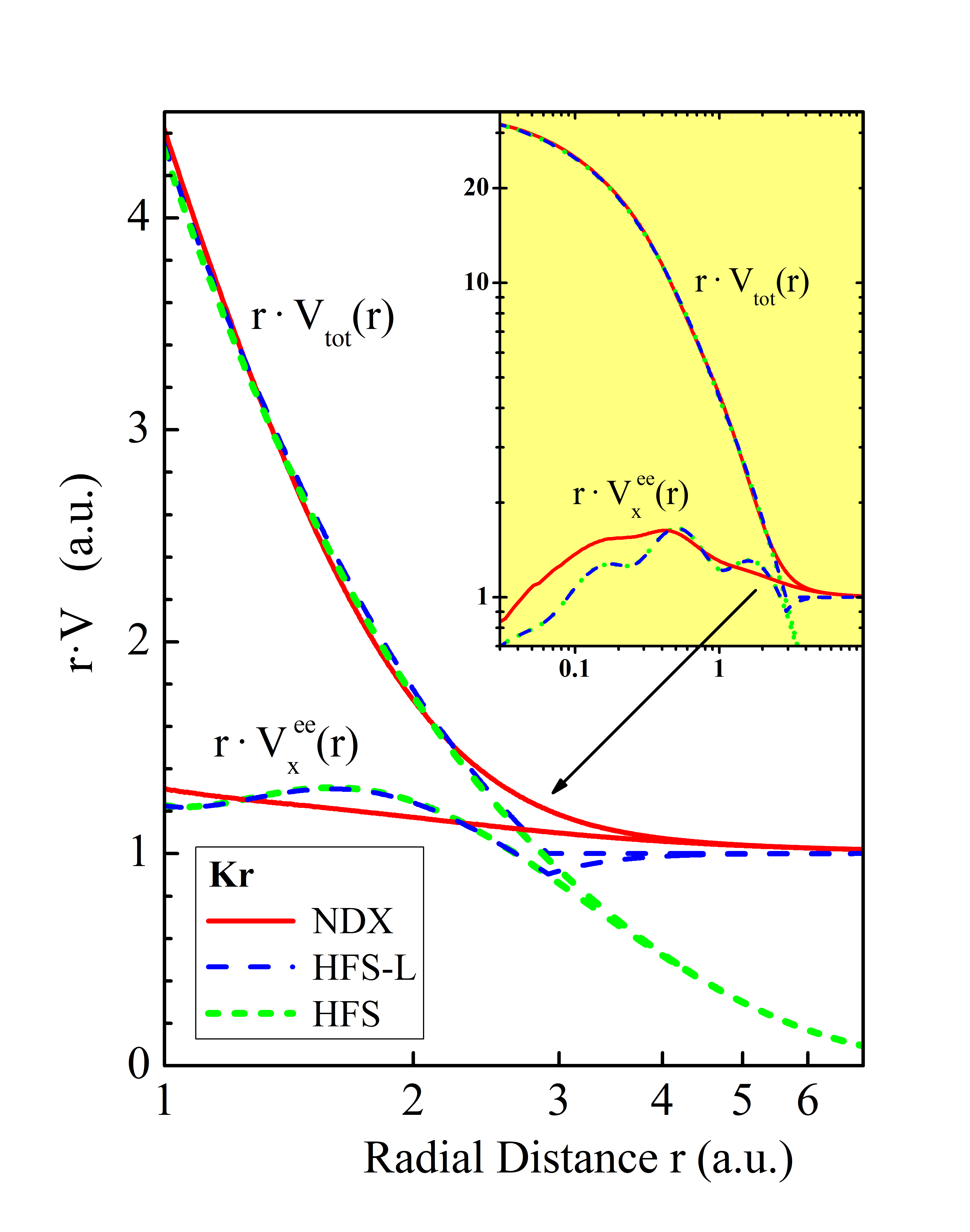}
    \caption{Scaled potentials (r$\cdot $V) for neutral Kr atoms
as function of the radial distance r. Shown are the total electronic
potential V$_{tot}$ as well as the corresponding exchange contribution V$%
_x^{e\,e}$ for different approximations: the present non-local approximation
to the Hartree-Fock exchange term (NDX, solid red curves), the
Hartree-Fock-Slater potential (HFS, short-dashed green curves) as well as the
Hartree-Fock-Slater potential including the Latter correction (HFS-L, dashed
blue curves). The inset shows the same results on extended logarithmic scales.}
    \label{fig:potKr}
\end{figure}

Fig. \ref{fig:potKr} displays scaled total potentials $r\,V_{tot}$ as well as
exchange potentials $r\,V_x^{e\,e}$ for fully converged self-consistent
fields (iterative solution of eq. \ref{HF}), calculated for the atomic
ground-state of Kr as function of the radial distance $r$. The results in
the plot are obtained from three different models:

\begin{itemize}
\item  the present NDX calculations (solid red curves).

\item  Hartree-Fock-Slater (HFS) calculations defined by eq. \ref{Slater},
as they are used, e.g. in solid-state electron dynamics (dotted green curves).

\item  Hartree-Fock-Slater calculations including the correction proposed by
Latter (dashed blue curves: HFS-L, see eq. \ref{Latt}). The corresponding
orbital energies calculated with our code show deviations from the
well-known Herman and Skillman calculations \cite{HermanSkillman} only at
the last digit given by these authors.
\end{itemize}

First one may note that the exchange contribution dominates the total
potential only at large radial distances $r$. The relative contribution of
the exchange potential decreases with decreasing $r$, contrary to the
absolute values of the exchange potential which has a maximum a $r=0$. HFS
and HFS-L results are very similar at small distances and they both show
bumps related to the orbital radii of the L, M and N shells. As discussed in
section \ref{subsec:HF}, $r\,V(r)$ has to approach $1$ at large values of $r$.
This is enforced by the Latter correction in the HFS-L model (see eq. \ref
{Latt}) and leads to the artificial kink in the potential near $r=3\,a.u.$ .
The standard HFS model fails completely in this region of $r$. But the NDX
potential approaches the correct asymptotic value and the calculations yield
much smoother exchange potentials compared with the other two models.
Furthermore, $V_{x,NDX}^{e\,e}$ exceeds the Slater FEG-exchange potential
(HFS and HFS-L) by 15 to 20\% at small $r$-values and should thus be close
to the true Hartree-Fock values \cite{HermanCallaway}.

\begin{figure}[htb!]
    \centering
    \includegraphics[width=0.55\linewidth]{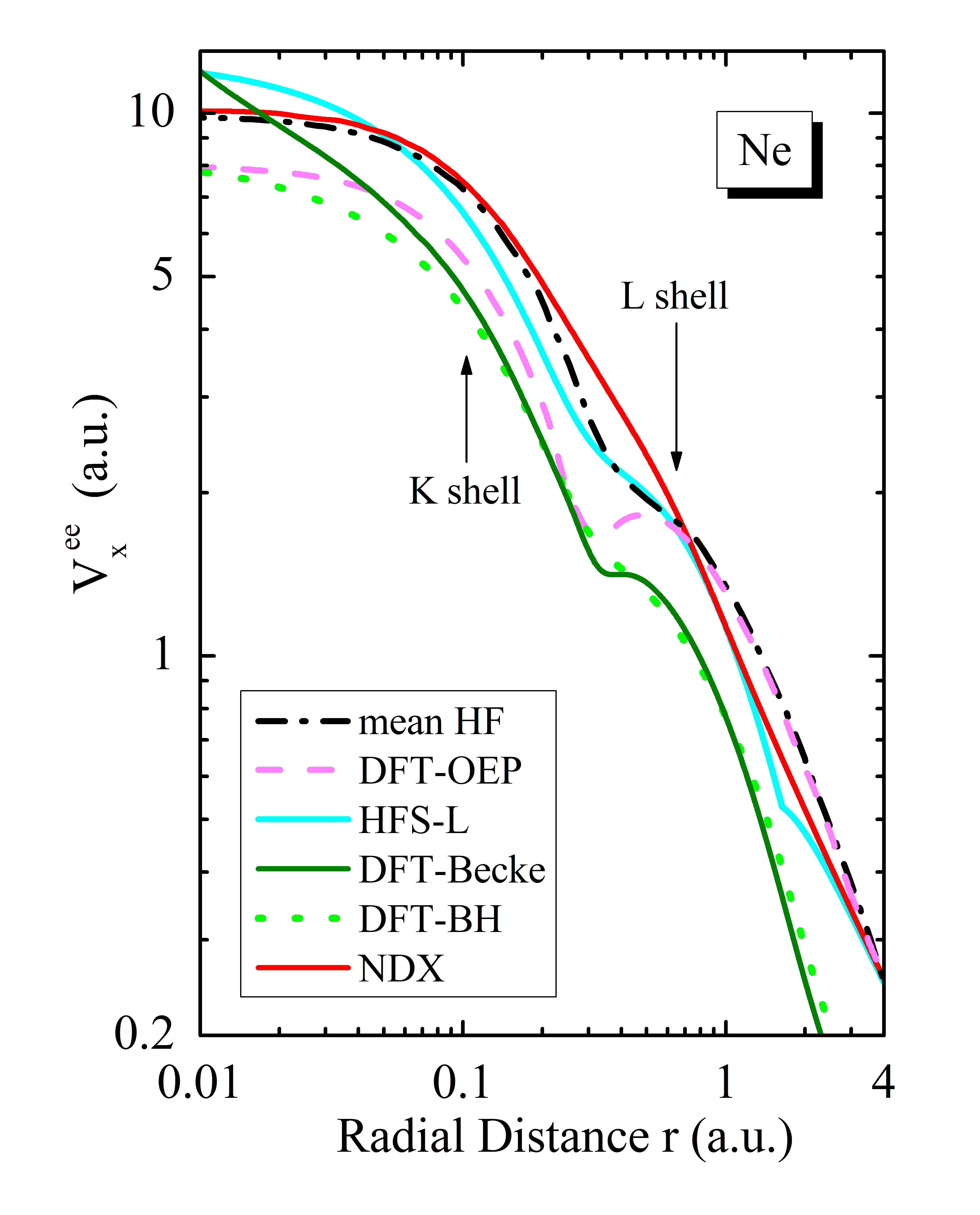}
    \caption{ Exchange potentials V$_x^{e\,e}$
for Ne plotted as function of the radial distance r. All results are based on 
SCF solutions for different approximations: the present non-local approximation to the Hartree-Fock
exchange term (NDX, solid red curve), the local potential method by von
Barth and Hedin (BH, dotted green curve) as well as the Hartree-Fock-Slater
potential including Latter correction (HFS-L, solid cyan curve). The
black dashed-dotted curve corresponds to an 
average non-local Hartree-Fock exchange term (mean HF) 
and the long-dashed magenta curve is a density-functional solution 
based on the optimized effective potential method (OEP). 
The solid olive curve uses the fitted exchange functional by Becke.}
    \label{fig:potNe}
\end{figure}

Fig. \ref{fig:potNe} displays exchange potential curves (V$_x^{e\,e}$) 
for different models as function of the radial distance in Ne.
As above all potentials correspond to fully converged self-consistent
fields. The red solid curve represents our NDX model and the solid cyan
curve stems from the HFS-L model and, of course, both models yield the
correct asymptotic exchange potential of $1/r$ for $r>4\,a.u$.
The DFT-Becke potential was also calculated with our program 
but using Becke's exchange functional \cite{Becke86}.
The other curves in the plot are taken from the very detailed work of Krieger et
al. \cite{Krieger}. The dotted green curve results from the local-spin
density method by von Barth and Hedin (BH) \cite{Barth}. 
Two of the results stem from solutions of the exchange potential, calculated with a
numerical effort that is comparable to the full Hartree-Fock solution. 
The black dashed-dotted curve is the {\em average} Hartree-Fock
potential calculated according to eq. \ref{aveX3}. The averaging procedure,
however, leads to significant deviations of the orbital eigenvalues from the
full Hartree-Fock solution (up to 22\% for the valence orbital of Zn) and to
deviations of the total energy \cite{Krieger}. 
The lower dashed curve corresponds to the so-called optimized
effective potential (OEP) method and is the exact non-local Kohn-Sham
solution without correlation \cite{Krieger,Talman}. 
Per definition, this model gives total energies that are the closest to the Hartree-Fock ground
state energy, if only a single potential is used for all occupied states of
a given spin direction. The corresponding OEP valence-orbital energies are
also nearly identical to the Hartree-Fock values. This extraordinary good
agreement, however, is paid with the price of large deviations for the other
orbital energies \cite{Krieger} 
related to the low $V_x^{e\,e}$ values around the K-shell position in Fig. \ref{fig:potNe}.

The NDX curve is seen to yield a rough representation of Slaters non-local
Hartree-Fock result (mean HF). Keeping the simplicity of the NDX model in
mind, this agreement looks promising. Only at intermediate distances 
below and above the L-shell radius there are significant deviations. At $r\approx 0.35\,a.u.$
all other models show indications of a dip in the potential due to the
reduced electron density in the space between the K and L shells. The NDX
potential does not show this dip, because it results from an average over
the surrounding (non-local) electron density without any destructive
interference terms, i.e., the NDX exchange hole contains no oscillatory
part (see eq. \ref{rhom-FEG}). Thus, it always tends to smooth 
all structures in the electron-density distribution. 

At $r=1.6\,a.u$ the HFS-L results in Fig. \ref{fig:potNe} show a similar kink as
discussed above for Kr. These HFS-L results incorporate Slaters local FEG exchange
term and are typically 10\% to 30\% lower than the accurate mean HF results
for all important regions of $r$. Only at distances very close to the
nucleus do the HFS-L results exceed the mean HF data. The local-spin density
results with the von Barth and Hedin exchange potential are nearly identical
to $2/3$ times the HFS-L data for $r<1.6\,a.u.$ since the BH exchange potential 
is consistent with the Kohn-Sham expression.
One may recognize that this approximation fails completely for large values
of $r$, similar as Becke's exchange approximation. 
These potentials do not follow the asymptotic $1/r$ behavior.

Many structural and dynamic problems properties, such as excitation, 
ionization or decay probabilities, are not only related to the wavefunctions, 
but also to the orbital energies. Thus, we feel that the deviations between the 
orbital eigenvalues of simplified models and full Hartree-Fock calculations 
need to be investigated. Since the $r$-dependence of
the potential is directly reflected in the orbital eigenenergies, these
energies as well as the total atomic energies will be investigated in the
next section.

\section{Atomic Energies}
\label{sec:Energies}

Since the r-dependence of exchange potentials are not easy to interpret,
it is most convenient to compare the energies calculated with different methods 
and with experimental data, where available.
In this work, we restrict ourselves to ground-state atoms from He to Xe.

\subsection{Orbital energies:}
\label{subsec:EOrbital}

The orbital eigenenergies $\varepsilon _\lambda $, as they follow from eq. 
\ref{HF}, depend significantly on the method used to calculate the exchange
potential. Here we compare atomic orbital energies from full relativistic
Dirac-Hartree-Fock (DHF) calculations (computed for a single ground-state 
configuration using the code of Grant et al. \cite{Grant}) with different simplified models.
To simplify the comparison, results of our NDX model, of HFS-L and of a DFT variant 
using the Latter restriction and Becke's exchange functional are plotted 
versus an effective quantum number n$_{eff}$ for all sub shells. 

\begin{figure}[htb!]
    \centering
    \includegraphics[width=0.6\linewidth]{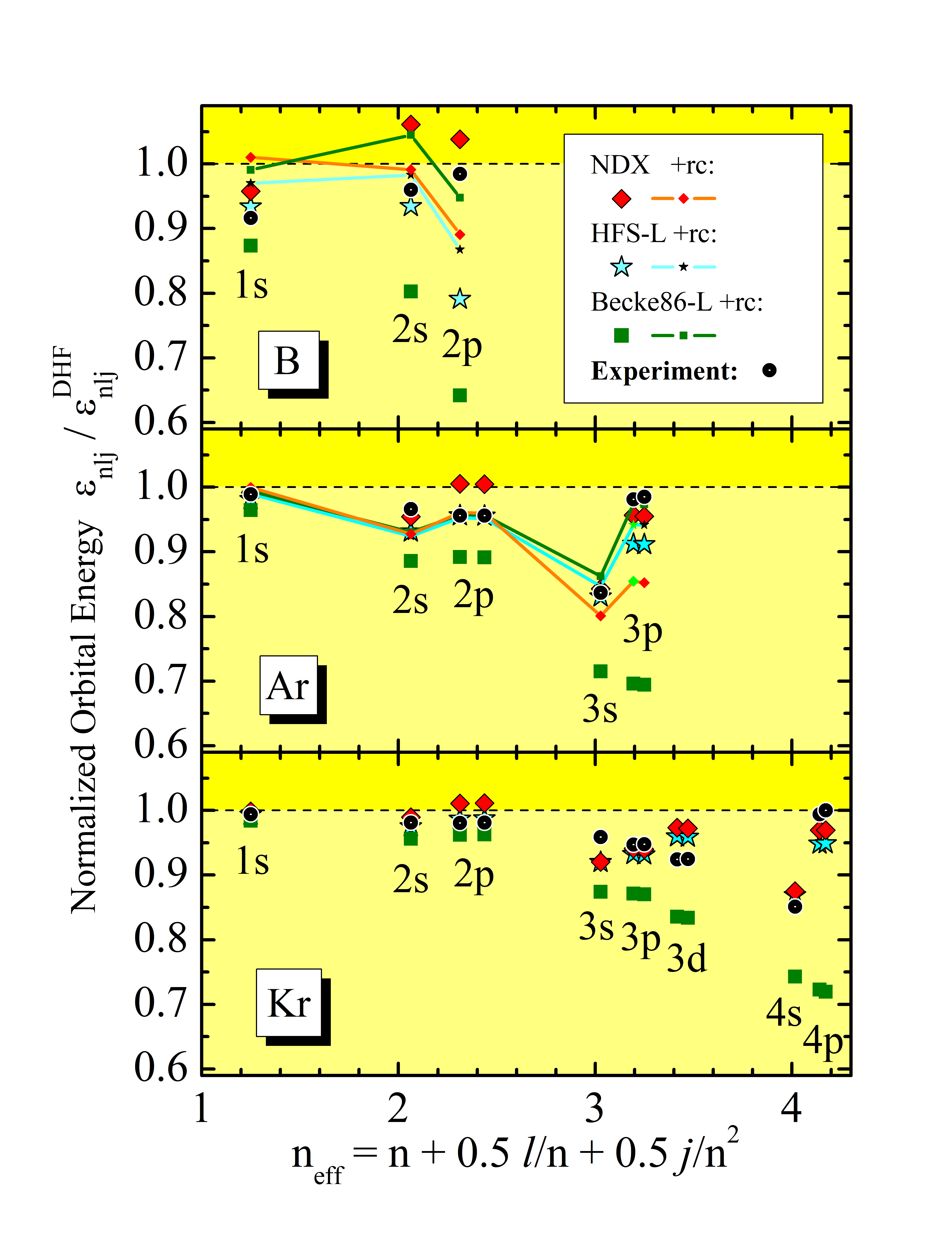}
    \caption{Orbital eigenenergies divided
by the ''exact'' Dirac-Hartree-Fock results for B (Z=5), Ar (Z=18) and Kr
(Z=36) atoms as function of an effective quantum number n$_{eff}$. Results
are shown for three different exchange approximations: the present NDX
model (red diamonds and orange lines), the HFS-L model (cyan asterisks and line) as well as
density funtional theory with Becke's exchange model accounting in addition 
for the Latter correction (solid olive squares and lines). 
Experimental orbital ionization energies are plotted as black circles. 
See appendix B for original DHF results and text for further details.}
    \label{fig:EOrbital}
\end{figure}

Fig. \ref{fig:EOrbital} displays the corresponding ratio of orbital energies from different
models to the exact DHF results $\varepsilon _{nlj} ^{DHF}$ for boron, argon and krypton. 
These energies and the corresponding self-consistent wavefunctions 
were calculated with the same computer code, which was previously 
applied to the prediction of ion energy losses in gases and solids 
\cite{NUM} using HFS-L potentials. For the numerical densities and potentials 
a linear grid of 10000 radial points has been used with a spatial boundary of 20 a.u.  
beyond which simple extrapolation is used. The estimated numerical
all-over accuracy is about 6 digits.

For the three atom types, the orbital energies are displayed by symbols 
and for B and Ar we have additionally determined orbital selective 
ionization potentials displayed by colored lines. The ionization potentials
are given by the difference of total energies $\Delta E_{nlj} = E^{tot}_{gs} - E^{tot}_{-nlj}$. 
Here, $E^{tot}_{gs}$ is the total energy (for details see next section) resulting from the 
corresponding ground-state SCF solution and $E^{tot}_{-nlj}$ indicates another SCF solution 
for the same atom, but with a vacancy in a specific orbital defined by the quantum numbers nlj. 
For all theoretical results in Fig. \ref{fig:EOrbital} as well as in the subsequent figures, 
we have included the following correction terms that allow for minor 
but straight forward improvements of the results to facilitate an easy comparison 
with existing experimental data (taken from refs. \cite{Moore, Bashkin}):
\begin{itemize}
\item We have considered the first-order relativistic mass--velocity, Darwin
and spin--orbit terms \cite{Foldy}, as they may be found in many textbooks 
\cite{Cowan,rSlater}. 

\item The Lamb shift was added as well.

\item The influence of the finite nuclear mass on the reduced electron mass 
as well as the deviation between finite nuclear size and the point charge in eq. \ref{Ham3}
(see, e.g., refs. \cite{Cowan,Grant,Bjorken}) lead to another small correction. 
\end{itemize}

For heavy atoms relativistic effects are the most important 
corrections to the total energy (after the dominant nucleus--electron interaction 
as well as the mean direct and exchange electron-electron interactions). The most important of the minor terms
for light atoms is the residual electron--electron interaction or correlation energy. 
As described in appendix A, we have applied a simple semiempirical correction 
to the total energies and correspondingly to the resulting ionization energies. 
Use of the sum of all correction terms, including relativistic and correlation corrections 
for the total energies is denoted by "+rc" in Fig. \ref{fig:EOrbital} 
(note that the correlation correction is not considered for the orbital energies).  

Condition (i) in the introduction seems to suggest that the ratios 
for the different orbital energies should be as close as possible to the ratio of one. 
The DHF results, however, are orbital energies and have been calculated for the 
atomic ground-state configuration only. Hence, they neither include relaxation nor correlation effects.
In other words, converged multi-configuration DHF total-energy calculations for the various initial and final states 
would be necessary to simulate the experimental results (black circles) at ratios between 0.83 and 1.0. 
This means, we judge about the accuracy of the considered models 
by comparison with the experimental data points. 

Fig. \ref{fig:EOrbital} indicates an improved agreement between all theories and 
experiment with increasing nuclear charge Z. Orbital energies determined 
with HFS-L and NDX show a similar and reasonable agreement with the data. 
Specifically for $\varepsilon _{valence}$ the NDX seems superior to HFS-L.
However, Becke's solution is far off, specifically for the outer shells. 
Additional DFT-Becke test calculations without the Latter restriction lead to wrong 
asymptotic potentials and consequently to even much lower valence-orbital energies. 
The ionization energies derived from the difference of total energies $\Delta E_{nlj}$
(depicted by solid lines) show a different behavior. For the B atom all three models yield 
similar deviations from experiment. For the inner shells of Ar the $\Delta E_{nlj}$ values 
of all three models do practically agree with each other and with the data. 
For the valence band of Ar Becke's results looks even perfect. A few selected $\Delta E_{nlj}$ 
calculations for Kr seem to show a similar trend as for Ar.

\begin{figure}[htb!]
    \centering
    \includegraphics[width=0.55\linewidth]{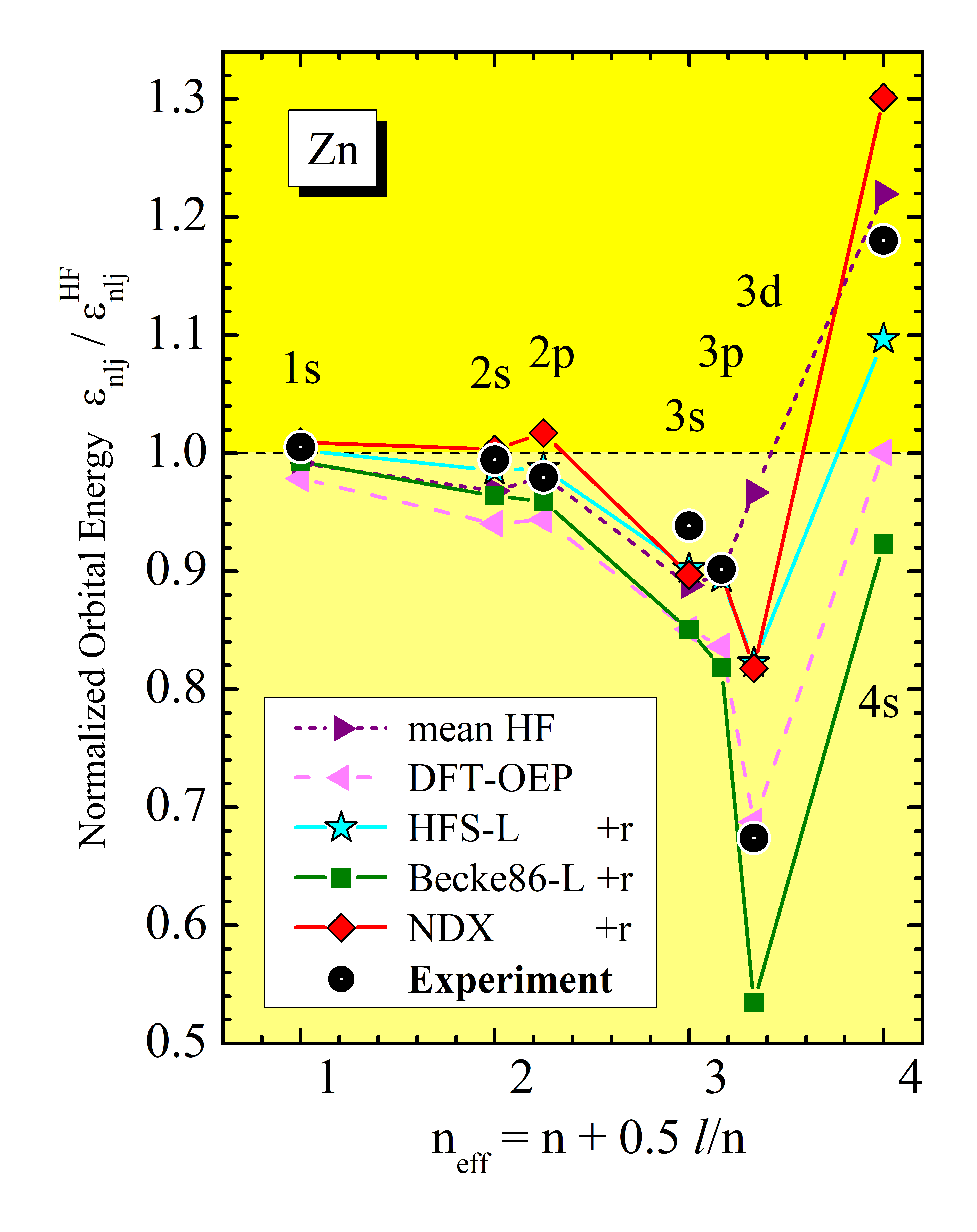}
    \caption{Orbital eigenenergies divided
by the ''exact'' non-relativistic Hartree-Fock results for atomic Zn (Z=30) 
as function of an effective quantum number n$_{eff}$. Results are shown for
five different exchange approximations: the present NDX model (solid red
diamonds), the HFS-L model (cyan asterisks), the DFT-Becke model with 
Latter correction (olive squares) as well as the DFT-OEP model
(magenta triangles) and the averaged Hartree-Fock exchange-model
(mean HF, purple triangles). 
For comparison experimental binding energies are shown as black circles. \cite{Moore, Bashkin}}
    \label{fig:ZnOrbitals}
\end{figure}

Fig. \ref{fig:ZnOrbitals} displays the ratio of orbital energies from different
models to non-relativistic Hartree-Fock (HF) results \cite{Froese} for zinc. 
The behavior of the HFS-L, DFT-Becke and NDX models relative to the 
experimental data is similar as discussed above with
the exception that the ratios for the valence-orbital energy are increased.
Judging from the larger deviation for 3s, 3p and 4s states DFT-Becke results 
involve the largest discrepancies from the data. Except for the 3d state 
DFT-OEP (uncorrelated Kohn-Sham solution taken from ref. \cite
{Krieger}) is not very much better and it involves by far 
the largest absolute deviation (about 200 eV for the 1s state).
As discussed above, the OEP valence-orbital
eigenvalue is nearly identical to the full HF value, 
reaching a ratio of about 1 (a typical DFT feature). 
The mean HF results (eq. \ref{aveX3}) are also 
taken from ref. \cite{Krieger}. With the exception of the 3d 
state these results are always close to the data.
This suggests that the mean HF potential leads to 
orbital eigenvalues that seem to incorporate relaxation effects.

Many models of dynamic electron properties need a single potential that
yields an accurate representation of the transition energies and thus, we
have also plotted the experimental electron binding-energies in Fig. 
\ref{fig:ZnOrbitals} \cite{Moore, Bashkin}. The averaged relative deviations of the
theoretical orbital energies from these binding energies are very similar
for all considered models. After correcting for relativistic effects, they range
between 7 and 12\%. It is known that correlation effects yield only a minor
influence on these deviations \cite{Cowan} and, as suggested by Koopmans's
theorem \cite{Koopmans}, the dominant contribution is a mean-field effect,
namely the relaxation of the residual orbitals after ionization. In order to
account for this relaxation effect correctly, theoretical binding energies
have to be calculated from the difference of the total energies of the
neutral and singly ionized systems, as shown in Fig. \ref{fig:EOrbital}.

\subsection{Total energies:}
\label{subsec:ETotal}

Multiplying the Schr\"{o}dinger equation (eq. \ref{Schr}) from the
right-hand side with the multi-electron wavefunction $\Psi _n$ for the
configuration $n$, we obtain the total energy $E_n^0$ for this
configuration. By using eqs. \ref{HF}, \ref{V1} and \ref{V2} and dropping the
configuration index for simplicity we have 
\begin{mathletters}
\begin{eqnarray}
E^0 &=&\left\langle H^0\right\rangle _\Psi  \label{tot1} \\
&=&\left\langle \sum_{i=1}^N\left[ T(\vec{r}_i)+V^{n\,e}(\vec{r}_i)\right]
+\sum_{i<j}^NV_{i,\,j}^{e\,e}(\vec{r}_i,\vec{r}_j)\right\rangle _\Psi
\label{tot2} \\
&=&\left\langle \sum_{i=1}^N\left[ T(\vec{r}_i)+V^{n\,e}(\vec{r}_i)\right]
+\sum_{i\neq j}^NV_{i,\,j}^{e\,e}(\vec{r}_i,\vec{r}_j)\right\rangle _\Psi
\label{tot3} \\
&&-0.5\sum_{i\neq j}^N\left\langle V_{i,\,j}^{e\,e}(\vec{r}_i,\vec{r}%
_j)\right\rangle _\Psi  \nonumber \\
&=&\sum_{\lambda =1}^N\left\langle T(\vec{r}_\lambda)+V^{n\,e}(\vec{r}%
_\lambda)+V_{direct}^{e\,e}(\vec{r}_\lambda)+V_{x,\lambda }^{e\,e}(\vec{r}%
_\lambda)\right\rangle _{\varphi _\lambda }  \label{tot4} \\
&&-0.5\sum_{\lambda =1}^N\left\langle V_{direct}^{e\,e}(\vec{r}%
_\lambda)+V_{x,\lambda }^{e\,e}(\vec{r}_\lambda)\right\rangle _{\varphi _\lambda } 
\nonumber \\
&=&\sum_{\lambda =1}^N\varepsilon _\lambda+0.5\sum_{\lambda =1}^N\left\langle V^{n\,e}(%
\vec{r}_\lambda)-V_{eff,\lambda }(\vec{r}_\lambda)\right\rangle _{\varphi _\lambda }  \
\label{tot5}
\end{eqnarray}
where $V_{eff,\lambda }$ are the total potentials (including the direct and
exchange electron--electron interaction terms as well as the interaction
with the nuclei) and $\varepsilon _\lambda $ are the orbital energies for
each state $\lambda $. Thus, the total electronic energy is equal to the sum
over the orbital energies plus a sum over simple terms that avoids double
counting of the electron--electron interaction energies.

$E^0$ as given by eq. \ref{tot5} is the exact total configuration-average
energy corresponding to eq. \ref{Ham3}, if the exact Hartree-Fock exchange
expression is used. So the eigenenergies $\varepsilon _\lambda $, the
wavefunctions and also the total effective potential $V_{eff}$ depend on the
model for the description of the exchange interaction. Note that many other
authors use the exact Hamiltonian $H^0$ and construct the Slater determinant
with the approximate wavefunctions \cite{Cowan,Krieger}. This usually gives
results that are very close to the Hartree-Fock value, but always higher
than Hartree-Fock as follows from the variational principle. Here we modify
the Hamilton operator by consistently keeping the different approximate
exchange terms for the HFS-L, DFT-Becke (including Latter correction) 
and NDX models in $V_{eff}$ of eq. \ref{tot5}.
This makes the results much more sensitive to the exchange approximation
than to the calculated wavefunction.

It is emphasized that the orbital energies  $\varepsilon _\lambda $  in  eq. \ref{tot5} 
include the same corrections as introduced already 
in section \ref{subsec:EOrbital}. For the total energy $E^0$ 
we have additionally considered the correlation correction 
according to the description in appendix A.
Hence, the all-over success of such models may 
be judged by comparison with the experimental total energies.

\begin{figure}[htb!]
    \centering
    \includegraphics[width=0.6\linewidth]{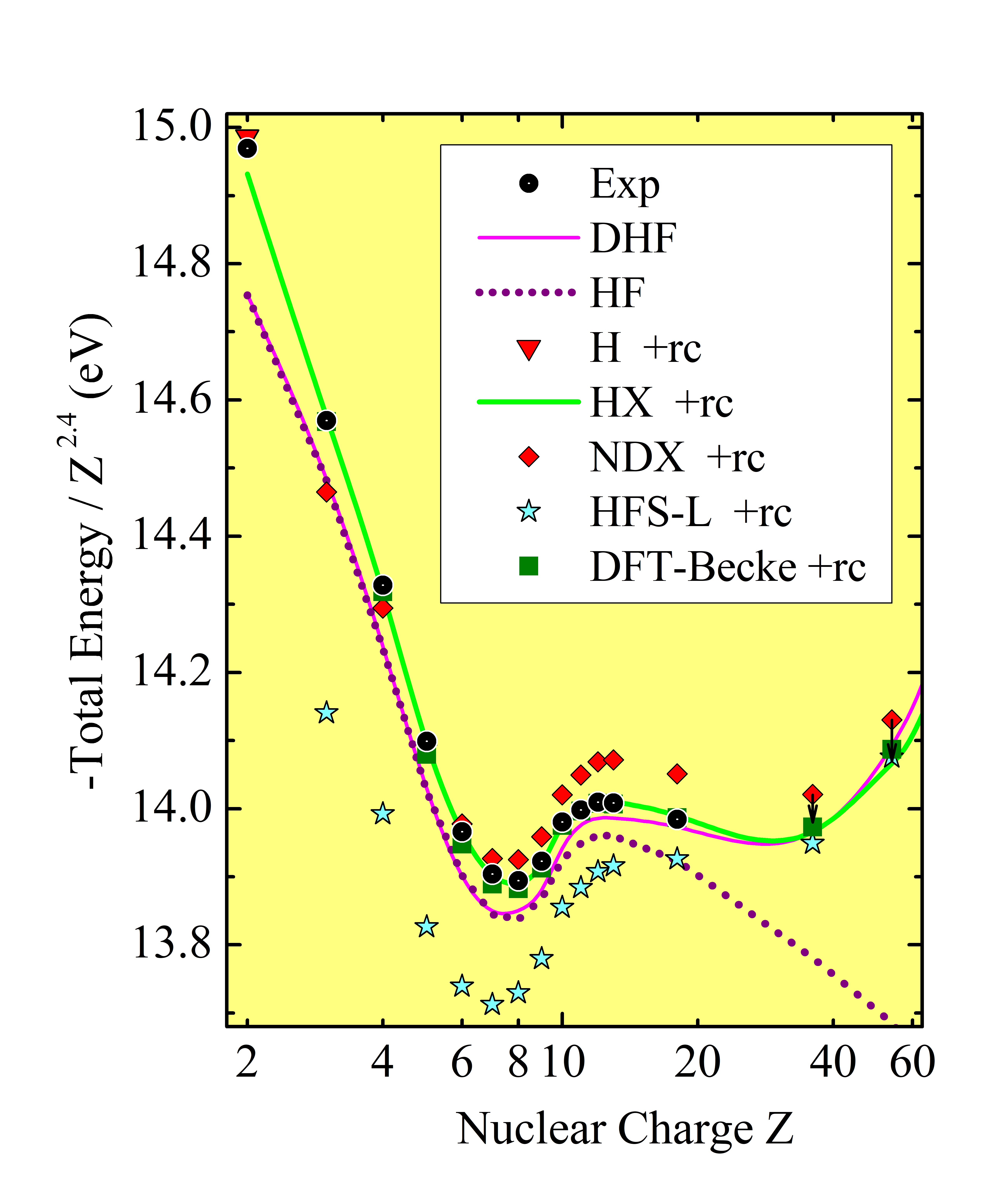}
    \caption{Total electronic
(configuration-average) energies for atoms from He to Xe as function of the
nuclear charge Z. Most of the theoretical data include correlation
corrections as well as relativistic corrections (+rc) to allow direct
comparison with the experimentally determined energies. The experimental
values (solid black circles) \protect\cite{Moore}, the Hartree-Fock values
(HF, dotted purple curve) \protect\cite{Blume} and the Hartree values with
statistical exchange (HX, solid green curve) are taken from the book by
Cowan \protect\cite{Cowan}. The full relativistic Dirac-Hartree-Fock values
(DHF, solid magenta curve) were calculated for a single ground-state
configuration using the program by Grant et al. \protect\cite{Grant}. The
present results (NDX) are shown as solid red diamonds. The HF result for He
(red triangle) as well as the HFS-L (cyan asterisks) and Becke's DFT results 
(solid olive squares) were also calculated with our SCF code.}
    \label{fig:ETotal}
\end{figure}

Fig. \ref{fig:ETotal} displays the total energy of neutral ground-state atoms as
function of the nuclear charge $Z$. The energies have been divided by $%
Z^{2.4}$ to provide a convenient graphical presentation. In the plot, two
sets of theoretical data (DHF and HF) without some specific corrections have
been included to show the influence of correlation and relativistic effects.
The full relativistic Dirac-Hartree-Fock (DHF) results are lower than the
experimental data for all nuclear charges. These DHF energies are calculated
for a single configuration (although multi-configuration calculations are
possible with the code by Grant et al. \cite{Grant}) and thus, they include
no correlation. Consequently, the difference between the DHF and
experimental energies is due to electron-correlation. The relative
contribution from electron-correlation is 1.5\% for He and reduces nearly
monotonically to 0.06\% for Ar. An estimate of the $Z$-dependence of the
correlation energy indicates that the plotted DHF results for $Z>50$ should
be very close to the exact results. The Hartree-Fock (HF) results for $Z\geq
2$ include no relativistic corrections and they are computed for the same
configurations as DHF. Thus, the deviations between the HF and DHF results
are due to relativistic effects. This contribution to the total energy
increases monotonically from 0.005\% for He to 3\% for Xe ($Z=54$).

All other theoretical results include correlation corrections as well as
relativistic corrections (+rc). Thus, they can directly be compared to the
experimental results. By far the best agreement with the experimental data
is found for the so-called HX results. The abbreviation HX stands for
''Hartree-plus-statistical-exchange'' method \cite{Cowan}. This method is an
approximation to the Hartree-Fock method (or an extended Hartree method),
which starts from replacing both sums $\sum_\mu $ in eqs. \ref{rho-d} and 
\ref{rho-x} by $\sum_{\mu \neq \lambda }$. Without the diagonal term $\mu
=\lambda $, the exchange potential $V_{x,\lambda }^{e\,e}(\vec{r})$ is
strongly reduced (there is no self interaction) compared to the HFS-L and
NDX models. Furthermore, the HX method is based on a statistical ansatz for
the corresponding exchange-charge density, plus a local-density exchange
approximation with empirically adjusted correction terms (depending on $\rho
(r)$, on the occupation numbers, on the orbital angular-momentum $\ell $ as
well as also on r) to fit exact Hartree-Fock results. The HX results
(including Cowan's specific relativistic and correlation corrections) may be taken as a
reliable estimate of the total experimental energies up to about $Z=33$. For 
$Z>36$ higher-order relativistic effects become important and the HX results
are lower than the DHF results. The main disadvantages of the HX model,
however, are the $\ell $-dependence of the potential and the resulting
non-orthogonal wavefunctions; the model contradicts conditions (ii), (iv)
and (vi) of the introduction. Hence, it is optimized for structural aspects
of the pure atomic case, but not applicable to molecular or solid-state
problems.

From the above discussion it follows that the NDX, HFS-L and Becke's results should
be compared to the existing experimental data up to $Z=18$, to the HX
results for $20\leq Z\leq 33$ and to the DHF values for $Z>50$. The HFS
results with Latter correction (HFS-L) are much too low at small nuclear charges and they slowly approach
the exact total energy. Contrarily, the NDX results seem to oscillate around
the exact results. For small Z, the corresponding deviations are about a
factor of six lower for NDX than the uncertainties of the HFS-L results.
Only at $Z\geq 20$ the HFS-L results seem to converge faster to the exact
energies.

This statement, however, turns out to be incorrect, if the influence of
higher-order relativistic effects for $Z>30$ is taken into account. For the
Kr and Xe case (Z=36 and 54) we have used an approximate scalar-relativistic
potential (similar potentials are discussed in refs. \cite{Cowan,rSlater})
to estimate the relativistic changes of the electron density. 
The corresponding self-consistent density was then frozen to calculate 
the resulting non-relativistic potentials, wavefunctions, energies and 
corrections (including the relativistic ones), as before. 
By doing this, we avoid including spurious 2$^{nd}$ order
relativistic effects (the scalar-relativistic potential includes the
relativistic effects only up to the 1$^{st}$ order) and may directly compare
to the results calculated for the non-relativistic SCF process.
For HFS-L and for NDX we compute very similar reductions of the
total energy which are indicated by the black arrows at the NDX values for
Kr and Xe. It is seen that the relativistic density effect brings the NDX values in
better agreement with DHF, whereas the HFS-L results move further below the reference values.
For Ar this relativistic density effect is clearly below the symbol size.
It follows that, except for the range of about $20\leq Z\leq 26$, the NDX energies are closer
to the exact total energy than the well known Hartree-Fock-Slater results
(even when including the Latter correction). 

The results of the calculations based on Becke's exchange expression \cite{Becke86}
(including Latter correction) are in very good agreement with the atomic total energy reference data. 
This agreement is nearly as good as found for the HX curve (see Fig. \ref{fig:ETotal}). 
One has to consider, however, that both are no ab-initio models 
(in contradiction to HFS-L and NDX). 
Becke's exchange as well as the HX exchange treatment are both based 
on fit parameter(s) that optimize total energy results for simple electronic ground-state systems. 
So, they might be much less accurate for complex atomic systems or 
for the description of excited states.
We have also calculated the HFS total energies without the Latter correction
(not shown in the figure) and the results are, dependent on the shell
structure, 0.5 to 1.5 eV below those with the Latter correction included.
Thus, especially for light atoms the HFS method, as it is often applied for
solid-state dynamics, yields larger deviations from the reference 
energies than all other results shown in Fig. \ref{fig:ETotal}. Furthermore, we
have also performed some test calculations with the Kohn-Sham exchange
expression ($\alpha =2/3$ in eq. \ref{Slater}) and the deviations from the
exact results increase by factors of 3 to 5 when compared to HFS-L, yielding
by far the worst agreement with the experimental data (or with Hartree-Fock,
if relativistic and correlation corrections are not included). For $Z<18$,
the local Kohn-Sham results would even not fit into the boundaries of Fig. 
\ref{fig:ETotal}. It is noted that the deviations between experiment and HFS,
HFS-L (with $\alpha =1$ and $2/3$) and also NDX is dominated by exchange effects.
Correlation effects, on the other hand, are only important for light atoms
such as He or Li (see Fig. \ref{fig:ETotal}). 

\section{Conclusions}
\label{sec:Conclusions}

Based on Hartree-Fock theory and on free-electron-gas results we propose a
new ab-initio method (NDX) for predicting exchange potentials in atoms, molecules,
clusters and solids. The improvement over previous local exchange potentials
is related to the direct use of an approximate extended exchange hole with
shapes depending on the distance $r$ from the nuclei. The resulting NDX
potential is unique for all electrons. Through the non-local density $\rho (%
\vec{r}^{\prime })$ it depends only on the coordinate vector $\vec{r}$ and,
unlike other simplified potentials, it fulfills all constraints that follow
from exact Hartree-Fock theory. Thus, the NDX potential should be
well-suited for application in density-functional calculations and may be
used for determinations of solid-state structure as well as in quantum
chemistry (see appendix C).

The average of this potential over all values of $r$ in a free electron-gas
is exactly the Slater exchange with the pre-factor $\alpha =1$. In the
regions close to the nuclei standard approximations (with $\alpha =2/3\,..\,1
$) underestimate the exchange potential significantly, whereas NDX
corresponds to a value of $\alpha \approx 1.3$ for the K-shell
electrons. For valence- and conduction-band electrons in solids it often
argued that the correct value of the exchange potential should be somewhere
between the Slater and Kohn-Sham ($\alpha =1$, respectively $\alpha =2/3$)
values. This requirement is automatically fulfilled with our method, since
delocalized electrons behave similar as plane waves and thus they test
mainly the interstitial space of solids, where the NDX potential corresponds
to $\alpha \approx 0.7$.

Furthermore, for individual atoms there are severe deviations between the
different potentials at distances exceeding about 1 \AA . It is known that
the common local-potential methods, e.g., those by Slater and Kohn-Sham,
fail completely for these large distances from the nucleus, since they
strongly underestimate the effects of exchange outside the regions of high
electron density \cite{Latter}. In solids, we expect an extraordinary strong
influence of this effect for large lattice spacings, e.g. in Van der Waals
aggregates, as well as for strongly localized valence-band states, as they
can be found in insulators and some semiconductors \cite{Gorling}. Within
the NDX method, however, this large-$r$ problem does not show up, since the
use of a properly normalized exchange hole yields the correct asymptotic
behavior of the exchange potential.

Comparison of self-consistent orbital energies for all shells and total energies,
calculated for atoms from He to Xe, clearly demonstrates that the present
NDX potential is superior to the Hartree-Fock-Slater (HFS) method or the
local Kohn-Sham potential (even when the large-$r$ failure of these
potentials is corrected). For light atoms (Li to Ne), the deviations from
experimental total energies are about a factor of seven lower for NDX than
for the well known HFS method. With respect to the uncertainties of orbital
energies the NDX method shows improvements over the HFS results by a factor
of two for the inner-most shells and by factors of up to 10 for the
outer-most shells. It is shown that the uncertainties of the HFS results for
outer shells of atoms can significantly be reduced if the so-called Latter
correction is applied. The NDX values, however, show generally a better
agreement with Dirac-Hartree-Fock (DHF) results or 
with experimental data if correlation corrections are considered.

The numerically much more involved mean Hartree-Fock method
(Slater average) as well as the optimized effective potential method (DFT solution)
yield orbital energies that do not come closer to the 
experimental binding energies than the simple HFS-L and NDX results. 
Cowan's HX method is not ab-initio and it is focused on atomic 
ground-state atoms. Thus, it cannot be applied to complex atomic structures. 
Becke's exchange approximation is also not ab-initio, 
but it is more flexibel and yields very good total atomic energies. 
This approach, however, yields the most significant deviations from 
Koopmans's theorem \cite{Koopmans}. In other words, the corresponding 
orbital energies are not in good agreement with the experimental data.

The NDX total energies deviate by only about 0.3\% from the experimental
values, even when the NDX exchange potential is used to compute the total
energies. A significant improvement may be reached when the NDX
wavefunctions are used to compute the Hartree-Fock energy expectation-value.
Approximate relativistic treatments and approximate correlation potentials,
however, are expected to limit the accuracy of all current models to about
0.1\% (except for atoms where much higher accuracies can be achieved for the
total energy). Exactly this uncertainty forces some solid-state structure
programs to make use of the orbital energies rather than the total energies.
Note that most electron-dynamics models are also based on orbital energies.
A possible solution for an application of the NDX method to multi-atomic 
systems is presented in appendix C. The success of the present NDX method in the
prediction of atomic energies suggests that it might be worth replacing current 
local-density approximations for the exchange potential in complex atomic systems.

\newpage
\section*{Appendix}

\subsection{Semiempirical correlation correction}
\label{subsec:Correlation}

The inclusion of correlation effects as described here is less firmly
founded than the above treatment of the exchange potential. Qualitatively,
correlation is the difference between the interaction of the smeared-out
charge clouds of all electron pairs (e$_i^{-}$,e$_j^{-}$) and the full
Coulombic electron--electron interaction $1/\,\left| \vec{r}_i-\vec{r}%
_j\right| $ of point charges (see the corresponding discussion in section 
\ref{subsec:HF}). With the full Coulomb interaction switched on, the electrons
mutually repel each other more strongly and the mean electron--electron
distances increase. Correspondingly, the repulsive electronic energies
decrease and the binding energies increase. Quantitatively, the correlation
energy is defined as the difference between the exact single-configuration
DHF results, including all the (minor) corrections for residual
single-electron effects, and the experimental total energy.

The effect of correlation has a non-local character, similar as discussed
for electron exchange in this work. Thus, it can be described by a displaced
correlation-charge density; a correlation hole. But since the correlation
potential is one to two orders of magnitude weaker than the local Slater
exchange potential (eq. \ref{Slater}), one may simply use a local
correlation correction without introducing significant uncertainties. Such
correlation corrections may be calculated most easily for the free electron
gas and accurate results are available for the cases of extreme low \cite
{Pines,CowanK,CarrC} and extreme high electron density \cite
{Gell-Mann,CarrM,Hedin}. Because it appears to be consistent with both
asymptotic free-electron-gas results, we have selected the following
expression for the correlation potential $V_{corr}$ by Gunnarsson \cite
{Gunnarsson} 
\end{mathletters}
\begin{equation}
V_{corr}(\vec{r})=-0.0225\,ln\left[ 1+21/r_s\left( \rho (\vec{r})\right)
\right] ,  \tag{A1}  \label{Vcorr0}
\end{equation}
where r$_s$ is a function of $\rho (\vec{r})$ given by eq. \ref{r-s}. In
this work, we have included an additional factor $\beta $ as a fitting
parameter to account for deviations from the free-electron-gas picture.
Furthermore, we have corrected this expression for the fact that the
self-interaction of electrons has to be excluded. Since the majority of all
electrons are localized near the nuclei, one may simply scale the electron
density from $N$ electrons per atom down to $N-1$ electrons and write 
\begin{equation}
V_{corr}(\vec{r})=-0.0225\;\beta \;ln\left[ 1+21/r_s\left( \rho (\vec{r})%
\frac{N-1}N\right) \right] .  \tag{A2}  \label{Vcorr}
\end{equation}
This density scaling yields a correlation potential of zero for hydrogen and
a significant suppression for He, but it introduces only minor changes for $%
Z\gg 3.$ The total correlation energy $E_{corr}$ is then 
\begin{equation}
E_{corr}=0.5\sum_{\lambda=1}^N\left\langle V_{corr}\left( \vec{r}_{\lambda}\right)
\right\rangle  \tag{A3}  \label{Ecorr}
\end{equation}
with a pre-factor of 0.5, to avoid double counting of the correlation energy
(each electron pair has to be counted only once). Note that the correlation
potential eq. A2 may also be added to the direct electron--electron
interaction terms in eq. \ref{V1}. But then one has to consider the double
counting of the potential electron--electron energy in exactly the same way
as for all other two-electron terms in eq. \ref{tot5}. The result of both
procedures, however, is nearly identical.

The parameter $\beta $ was determined without making any use of the current
NDX model. First we have performed Hartree-Fock-Slater calculations
including the Latter correction and with $\alpha $ in eq. \ref{Slater} taken
as a free parameter (so-called X$_\alpha $ calculations). We have performed
the corresponding X$_\alpha $ fits for atoms from Li (with $\alpha =1.1785$)
to Ar (with $\alpha =1.088$) and with $\alpha \approx 1+0.28/Z^{0.4}$ the
corresponding total energy without any correction agrees nearly perfectly
with exact Hartree-Fock results \cite{Blume,Cowan}. With the resulting
wavefunctions, we have applied all corrections discussed above 
including the correlation correction. The parameter $\beta $ was then
fixed by a fit to the experimental total energies \cite{Moore}. Thus, the
correlation correction is optimized for an X$_\alpha $ variant of the HFS-L
model in Fig. \ref{fig:ETotal} and the average over all results for 11 atoms
from He (with $\beta =0.689$) to Ar (with $\beta =0.73$) is 
\begin{equation}
\bar{\beta}=0.74\pm 0.10.  \tag{A4}  \label{BetaCorr}
\end{equation}
This low value points to a suppression of electron correlation in atoms
compared to a free electron gas. In fact, one may speculate that the
attractive nucleus restores radial electronic motions, which virtually would
reduce the number of degrees of freedom from 3 (one radial plus two angular
degrees of freedom) to 2, corresponding to $\beta =2/3$. 
Another argument for a mean value of $\beta$ below 1 
is the atomic shell structure itself. Typically, electrons interact significantly 
only with a few other electrons in the same shell or subshell. 
For s electrons there are spatial regions where 50\% of the density 
should count as a self interaction and nearly 100\% for valence electrons in alkali metals.
In this work, we have used eqs. A2 and
A3 with $\bar{\beta}=0.74$ for the HFS-L, DFT-Becke (plus Latter correction) 
and the NDX results for the total energies.

\newpage
\subsection{Used HF and DHF reference values}
\label{subsec:reference}

For reference purposes and for enabling to trace back the plotted curves 
in Figs. \ref{fig:EOrbital} and \ref{fig:ZnOrbitals}
to the original orbital energies and orbital binding energies,
we show the corresponding Dirac-Hartree-Fock (DHF) and 
non-relativistic Hartree-Fock results in the following two tables:

\begin{table}[htb!]
    \centering
    \begin{tabular}{llllr}
    \textbf{Atom} & \textbf{n  } & \textbf{l  } & \textbf{j       } & \textbf{$\varepsilon _\lambda (a.u.)$  } \\
    B  & 1 & 0 & 0.5 & 7.69754  \\
    B  & 2 & 0 & 0.5 & 0.49488  \\
    B  & 2 & 1 & 0.5 & 0.30982  \\
    --- &   &   &   &    \\
    Ar& 1 & 0 & 0.5 &  119.12662  \\
    Ar& 2 & 0 & 0.5 &   12.41158 \\
    Ar& 2 & 1 & 0.5 &   9.63196 \\
    Ar& 2 & 1 & 1.5 &   9.54707 \\
    Ar& 3 & 0 & 0.5 &   1.28658 \\
    Ar& 3 & 1 & 0.5 &   0.59539 \\
    Ar& 3 & 1 & 1.5 &   0.58782 \\
    --- &   &   &   &    \\
    Kr& 1 & 0 & 0.5 & 529.68545  \\
    Kr& 2 & 0 & 0.5 & 72.07981  \\
    Kr& 2 & 1 & 0.5 & 64.87475  \\
    Kr& 2 & 1 & 1.5 & 62.87923  \\
    Kr& 3 & 0 & 0.5 & 11.22445  \\
    Kr& 3 & 1 & 0.5 & 8.61987  \\
    Kr& 3 & 1 & 1.5 & 8.31281  \\
    Kr& 3 & 2 & 1.5 & 3.77766  \\
    Kr& 3 & 2 & 2.5 & 3.72681  \\
    Kr& 4 & 0 & 0.5 & 1.18774  \\
    Kr& 4 & 1 & 0.5 & 0.54151  \\
    Kr& 4 & 1 & 1.5 & 0.51435
    \end{tabular}
    \caption{''Exact'' Dirac-Hartree-Fock single-configuration results for neutral B (Z=5), Ar (Z=18) 
and Kr (Z=36) atoms are displayed in atomic units as function of the quantum numbers. 
We have computed these values using the full relativistic DHF code by Grant et al. \cite{Grant}}
    \label{tab:bessy_params}
\end{table}

\begin{table}[htb!]
    \centering
    \begin{tabular}{lllr}
    \textbf{Atom} & \textbf{n  } & \textbf{l  } &  \textbf{$\varepsilon _\lambda (Ry)$} \\
    Zn & 1 & 0 & 706.609  \\
    Zn & 2 & 0 & 88.723  \\
    Zn & 2 & 1 & 77.850  \\
    Zn & 3 & 0 & 11.276  \\
    Zn & 3 & 1 & 7.679  \\
    Zn & 3 & 2 & 1.565  \\
    Zn & 4 & 0 & 0.585
    \end{tabular}
    \caption{''Exact'' non-relativistic Hartree-Fock single-configuration results for neutral Zn atoms (Z=30)
 are displayed in Rydberg units (= 0.5 a.u.) as function of the quantum numbers. 
The values have been taken from a publication by Charlotte Froese-Fischer \cite{Froese}}
    \label{tab:bessy_params}
\end{table}

\newpage
\subsection{NDX for multi-atomic systems}
\label{subsec:solids}

For molecules, clusters, bulk solids or surfaces eqs. \ref{G7}, \ref{G8} and \ref{G9}
may directly be used for the computation of the NDX exchange potential $V_{x,NDX}^{e\,e}(\vec {r})$. 
It is clear, however, that the integral NDX method is computationally 
more challenging than differential methods.
The parallel integration of the four basic terms discussed in relation to eqs. \ref{G8} and \ref{G9} 
should also be an advantage for more complex systems. 
For some systems, consideration of symmetry properties 
can reduce the computational efforts. Furthermore, an exclusive 
restriction to orbital energies should significantly reduce the accuracy requirements and computing time.

The extension of the NDX to multi-atomic or even multi-elemental systems is not 
consistent with the above atomic definition of the parameter $\alpha$. 
The corresponding eqs. \ref{G10} and \ref{G11} consider only a single nucleus of charge Z.
Therefore, we suggest a generalization of these equations that is numerically 
relatively easy to handle and follows the same idea. Electron coordinates far away from 
all atoms or in interstitial regions are most important for weakly bound electronic states 
such as surface states, valence- and conduction-band  states.
Such coordinates $\vec r$ (at larger distances from all nuclei of the system) 
should lead to the Kohn-Sham solution $\alpha \approx 2/3$. 
Deeply bound electrons, however, are close to one nucleus and thus, 
coordinates in the vicinity of any of the nuclei belong to the other extreme,
namely $\alpha \approx 4/3$.

For multi-atomic systems, we use nearly the same distance definition as 
in eq. \ref{G11} for a single atom. 
Here the fractional electron charge $Q( \vec{R_i}, \vec{r} ) /Z_i$ inside a sphere of radius $r_{max,i} = \left| \vec{R_i} - \vec{r} \, \right|$ 
around the nucleus with index i is given below (consideration of the shape of
Wigner-Seitz cell boundaries for each atom seems not to be necessary)
\begin{equation}
f^{far} \left( \vec{r}, i \right) = Q \left( \vec{R_i}, \vec{r} \right) /Z_i = \frac{1}{Z_i} \cdot \int_{r \prime \, \leq  \, r_{max,i}} d\vec{r} \, \prime \;
\rho \left( \vec{r} \, \prime \right).  \tag{C1} \label{G11solid}
\end{equation}
The function $f^{far} \left( \vec{r}, i \right) $ uses the fractional charge 
as a density dependent measure of the distance from nucleus i. 
The involved numerical integration may be truncated if the integral reaches or exceeds $Z_i$.
This function is zero close to the nucleus i (for $\vec{r} \approx \vec{R_i}$) and 
reaches large values for electron coordinates far away from the nucleus. 
Values close to or above 1 could in principle mean that the coordinate $\vec{r}$ belongs to a conduction 
or valence band. However, the same coordinate might also be related to a small value 
$f^{far} \left( \vec{r}, j \right)$ at another atom with index $j \neq i$. 
Under this condition, the coordinate belongs to a bound state at atom j 
and not to any of the valence or conduction bands.
This problem is accounted for within the following linear scaling for $\alpha $ 
\begin{equation}
\alpha (\vec{r}) = 1.298 - 2\cdot 0.298\cdot min[1, f^{far} \left( \vec{r}, i \right) , f^{far} \left( \vec{r}, i+1 \right) , ...] 
\tag{C2} \label{G10solid}
\end{equation}
For each of the SCF iterations $\alpha (\vec{r})$ has to be computed for all values of $\vec{r}$. 
The $min$ function is given solely by the smallest fractional-charge distance 
(related to one of the nuclei close to the considered electron coordinate).
Thus, it is suggested to determine a few fixed nearest neighbor positions around each electron coordinate $\vec{r}$ 
and store the corresponding atomic indices that are to be evaluated 
within eq. \ref{G10solid} (the scaling function for $\alpha (\vec{r})$). 
An average of eq. \ref{G10solid} over the atomic charge distribution 
should yield the Slater result $\bar{\alpha}=1$ for a homo-nuclear system.
This means it is consistent with condition (iv) of the introduction (see p. \pageref{conditions}).

\section*{Acknowledgment}
We are very much indebted to H. Haas for many suggestions 
on a previous version of this manuscript.

\end{document}